\documentclass[letterpaper,twocolumn,10pt]{article}

\PassOptionsToPackage{dvipsnames,table}{xcolor}
\usepackage{usenix-2020-09}

\usepackage{prp-macros}
\usepackage{mawa-macros}
\usepackage{ow-macros}
\usepackage{bj-macros}

\usepackage{cleveref}
\crefname{figure}{Fig.}{Figs.}
\Crefname{figure}{Fig.}{Figs.}
\crefname{section}{\S}{\S\S}
\Crefname{section}{\S}{\S\S}
\newcommand{\crefsectionformat}[1]{%
    \crefformat{#1}{\S##2##1##3}%
    \Crefformat{#1}{\S##2##1##3}%
    \crefrangeformat{#1}{\S\S##3##1##4\crefrangeconjunction##5##2##6}%
    \Crefrangeformat{#1}{\S\S##3##1##4\crefrangeconjunction##5##2##6}%
    \crefmultiformat{#1}{\S\S##2##1##3}{\crefpairconjunction##2##1##3}{\crefmiddleconjunction##2##1##3}{\creflastconjunction##2##1##3}%
    \Crefmultiformat{#1}{\S\S##2##1##3}{\crefpairconjunction##2##1##3}{\crefmiddleconjunction##2##1##3}{\creflastconjunction##2##1##3}%
    \crefrangemultiformat{#1}{\S\S##3##1##4\crefrangeconjunction##5##2##6}{\crefpairconjunction##3##1##4\crefrangeconjunction##5##2##6}{\crefmiddleconjunction##3##1##4\crefrangeconjunction##5##2##6}{\creflastconjunction##3##1##4\crefrangeconjunction##5##2##6}%
    \Crefrangemultiformat{#1}{\S\S##3##1##4\crefrangeconjunction##5##2##6}{\crefpairconjunction##3##1##4\crefrangeconjunction##5##2##6}{\crefmiddleconjunction##3##1##4\crefrangeconjunction##5##2##6}{\creflastconjunction##3##1##4\crefrangeconjunction##5##2##6}%
}
\crefsectionformat{section}
\crefsectionformat{subsection}
\crefsectionformat{subsubsection}

\newcommand{\sys}{\textsc{Scepsy}\xspace}
\newcommand{\alp}[1]{Aggregate LLM Pipeline\xspace}
\newcommand{\alpacr}[1]{ALP\xspace}
\newcommand{\gridpanel}[1]{\includegraphics[width=\linewidth,trim=8 9 9 9,clip]{#1}}

\makeatletter
\def\section{\@startsection{section}{1}{\z@}{-3.5ex plus -1ex minus -.2ex}{1.2ex plus .2ex}{\reset@font\large\bf}}
\def\subsection{\@startsection{subsection}{2}{\z@}{-3.25ex plus -1ex minus -.2ex}{0.8ex plus .2ex}{\normalfont\large\bfseries}}
\makeatother

\begin{document}

\date{}

\title{\sys{}: Serving Agentic Workflows Using Aggregate LLM Pipelines}

\author{
  \rm
    Otto White$^{*\dagger}$ \quad
    Marcel Wagenl{\"a}nder$^{*\dagger}$ \quad
    Britannio Jarrett$^{\ddagger}$ \quad
    Xijin Zhou $^{\dagger}$ \quad
    Yanda Tao$^{\dagger}$ \\
\rm
    Pedro Silvestre$^{\dagger}$ \quad
    Guo Li$^{\dagger}$ \quad
    Huanzhou Zhu$^{\dagger}$ \quad
    Llu\'is Vilanova$^{\dagger}$ \quad
    Peter Pietzuch$^{\dagger}$ \\[4pt]
  {\small
    $^{*}$Equal contribution \quad
    $^{\dagger}$Imperial College London \quad
    $^{\ddagger}$Independent Researcher
  }
}

\maketitle

\pagestyle{plain}

\begin{abstract}
Agentic workflows carry out complex tasks by orchestrating multiple large language models~(LLMs) and tools. Serving them at a target throughput with low latency is hard because they are written in arbitrary agentic frameworks and their execution times are unpredictable: execution branches, fans out, or recurs in data-dependent ways. Since their LLMs often outnumber the available GPUs, they also oversubscribe GPUs.

We describe \sys{}, a serving system that schedules arbitrary multi-LLM agentic workflows onto a GPU cluster. \sys{} exploits the insight that, while the end-to-end latency of an agentic workflow is unpredictable, each LLM's fraction of execution time is comparatively stable across requests. \sys{} profiles each LLM under different parallelism degrees and combines the profiles with these fractions into an \alp{}, a lightweight throughput and latency predictor for allocations. To minimize latency at a target throughput, \sys{} uses the \alp{} to search over fractional GPU shares, tensor parallelism degrees, and replica counts. A hierarchical heuristic then places the chosen allocation onto the cluster, minimizing fragmentation and respecting network topology. On realistic agentic workflows, \sys{} achieves up to $2.5\times$ higher throughput before saturation and $1.0$--$3.3\times$ lower latency than systems that optimize LLMs independently or rely on user-specified allocations.
\end{abstract}

\section{Introduction}
\label{sec:intro}

Large language models~(LLMs) can solve complex problems and generate high-quality text~\cite{chatgpt, claude, gemini}, and rely on parallelization over multiple GPUs to provide fast and cost-effective responses.
Yet individual LLMs are insufficient for complex tasks due to finite context windows~\cite{attn-is-all-you-need}, lack of external or domain-specific knowledge~\cite{rag}, and increasing hallucinations during extended reasoning~\cite{cot}.

As a result, users increasingly build \emph{agentic workflows} that orchestrate multiple LLMs and tools to enable new capabilities and improve accuracy-cost trade-offs~\cite{frugalgpt}. Coding agents are a prime example of this~\cite{opencode}, using automatic model routers~\cite{openrouter, copilot-auto} to pick an LLM per prompt based on the task, or prompt libraries~\cite{superpowers-sdd} to instruct coding agents to run mechanical edits on cheap models and complex tasks on more capable ones. Workflows may also leverage multi-model inference-time scaling~\cite{test-time-scaling} (\eg beam~\cite{beam-with-prm} and Monte Carlo tree search~\cite{mcts}), or multi-model document processing and retrieval pipelines~\cite{rag-reranker, flashrag} to enable structured reasoning.

Hyperscalers have large data centers and performance engineering teams to optimize the serving of such agentic workflows~\cite{claude,gemini,chatgpt}, but smaller organizations typically have less expertise and fewer resources to self-manage deployments efficiently, and in small-to-medium-scale GPU clusters even small inefficiencies matter. Yet self-managed deployments of agentic workflows are still desirable for reasons of privacy~\cite{llm-privacy} and cost savings~\cite{aiindexreport}. We identify four challenges in self-managed deployments of complex agentic workloads:

\tinyskip

\mypar{Arbitrary agentic workflows and frameworks} Users deploy a variety of agentic frameworks, such as LangChain~\cite{langchain,langgraph}, AutoGen~\cite{autogen}, or Camel~\cite{camel}, and other libraries (schema validation, numerical, databases) to define workflows~\cite{agent-programming-models}. Thus, a serving system cannot assume a single workflow representation, programming model, or API.

\mypar{Unpredictable workflow execution} The latency of LLM execution is highly variable, because generation proceeds token-by-token until the end of the sequence. Since LLMs may invoke tools and other LLMs, the workflow itself may branch, iterate, and recur in data-dependent ways. These properties make predicting the throughput and latency of an agentic workflow based on a given GPU allocation challenging.

\mypar{Conflicting optimization objectives} Users want agentic workflows with high throughput and low latency. LLM throughput can be increased by exploiting \emph{data parallelism} with parallel model replicas, but this reduces the GPU resources available for \emph{tensor parallelism}, which lowers latency. While serving a single LLM navigates one such trade-off~\cite{alpaserve, kubernetes}, agentic workflows with multiple LLMs compose several objectives: adjusting the allocation of one LLM may affect others, potentially shifting bottlenecks. The heterogeneity of GPU interconnects worsens this challenge~\cite{nvlink}, because it constrains tensor parallelism.

\mypar{Oversubscribed GPUs} GPUs are a scarce and expensive resource~\cite{aiindexreport}, especially in small-to-medium-scale clusters. Since agentic workflows have a highly heterogeneous mix of LLMs (embedding, generator, reward, etc.), coarse-grained LLM-to-GPU allocations can lead to poor GPU utilization.

\tinyskip

Current solutions fail to address these challenges. Some serving approaches for agentic workflows exploit the workflow structure to optimize execution, \eg through graph-aware batching, KV cache sharing, or reduced head-of-line blocking~\cite{parrot, tokencake, agentix}. Several of these approaches~\cite{parrot, tokencake}, however, are constrained by framework-specific programming models, and most are limited to single-LLM workflows. More fundamentally, they focus solely on temporal request scheduling to GPUs, and leave the allocation of LLMs to GPUs to users. Some approaches~\cite{alpaserve, muxserve, prism, kubernetes, aegaeon} focus on multi-LLM serving through temporal or spatial multiplexing without assuming any specific agentic workflow structure. This makes them more flexible, but the lack of workflow awareness reduces performance, because LLMs are scheduled in isolation.

We make the observation that, whereas LLM execution time is hard to predict in agentic workflows, the fraction of time spent executing on each LLM is relatively stable across workflow requests. As we show in \cref{subsec:limitations}, in a beam search agentic workflow, the ratio between the generator and verifier LLMs is structurally fixed: each generator LLM invocation is followed by a verifier LLM invocation, so a slow request in one is also slow in the other. In a coding agent, the larger orchestration LLM's share of time is also fixed even when it launches a varying number of sub-agents on a smaller LLM\@.

Our insight is that a scheduler does not need to model a workflow's control flow to allocate GPUs, but requires the \emph{aggregate} demand that each LLM receives per workflow request. Under aggregate demand, a workflow behaves as a \emph{pipeline}, with negligible non-LLM tool and orchestration time (at most a few milliseconds in our measurements). The LLM with the least spare capacity thus bounds throughput, whereas end-to-end latency is the sum of the per-LLM latencies, adjusted for the average invocation concurrency. The fractions hold across requests, but not necessarily across time: when arrival rates or workflow parameters such as the beam width change, an allocation needs updating to account for the changes in how often a workflow request invokes each LLM\@.

We describe \sys{}, an agentic workflow serving system that efficiently schedules LLMs onto GPUs according to their relative performance contributions. \sys{} makes the following new technical contributions:

\mypar{(1)~Modeling agentic workflows as Aggregate LLM Pipelines~(\S\ref{sec:performance-model})} \sys{} makes GPU allocation decisions based on \emph{aggregate} per-LLM statistics collected over many executions. To support arbitrary agentic workflow frameworks~\cite{langchain, langgraph, autogen, camel}, \sys{} extracts these statistics through a combination of tracing low-level LLM invocations and LLM performance profiling, each without knowledge of the workflows or frameworks themselves. It then leverages these statistics to construct an \alp{}, a simplified representation of the resources used by the workflow, which enables pipeline-style latency and throughput predictions of agentic workflows under different GPU allocations.

\mypar{(2)~Joint throughput/latency optimization of GPU allocations~(\S\ref{sec:resource-scheduling})} Agentic workflows combine LLMs with different resource needs, which are coupled through workflows. \sys{}'s scheduler therefore performs joint, throughput- and latency-aware GPU allocation based on a novel search space that considers model \emph{parallelism} together with \emph{fractional} GPU allocations: for each LLM, \sys{} chooses (i)~how many model replicas it deploys; (ii)~what tensor parallelism degree each replica uses; and (iii)~what fraction of a GPU is given to each replica. Fractional GPU allocation allows \sys{}'s scheduler to maximize GPU utilization, ensuring model replicas are not overdimensioned by modeling GPUs as quasi-continuous resources. Considering these three dimensions is sufficient to optimize the throughput/latency trade-off: model replicas add serving capacity to bottlenecked LLMs, tensor parallelism reduces latency for LLMs on the critical path, and fractional GPU allocation increases GPU utilization.

The same search also adapts a deployed allocation when the workload shifts: live execution traces update arrival rates and invocation statistics for the \alp{}, while the LLM profiles remain valid to reason about a continuous performance space. \sys{} then reconfigures the cluster only when the predicted gain justifies it.

\definecolor{tickgreen}{RGB}{34,139,34}
\definecolor{crossred}{RGB}{200,50,50}
\newcommand{\goodtick}{{\color{tickgreen}\ding{51}}}
\newcommand{\badcross}{{\color{crossred}\ding{55}}}
\definecolor{lightgray}{RGB}{220,220,220}
\newcommand{\lightrule}{\arrayrulecolor{lightgray}\cmidrule{3-10}\arrayrulecolor{black}}
\newcommand{\subgroup}[3]{\multirow{#2}{*}{\rotatebox[origin=c]{90}{\parbox{#1}{\scriptsize\textbf{#3}}}}}

\begin{table*}[t]
\centering
\footnotesize
\setlength{\aboverulesep}{0pt}
\setlength{\belowrulesep}{0pt}
\renewcommand{\arraystretch}{1.0}
\setlength{\heavyrulewidth}{1.2pt}
\setlength{\tabcolsep}{4.5pt}
\begin{tabular}{@{}cll!{\vrule width 1.2pt}clccccc@{}}
\toprule
& & \textbf{System} & \shortstack[c]{\rule{0pt}{1.2em}\textbf{Optimization}\\\textbf{target}} & \shortstack[c]{\rule{0pt}{1.2em}\textbf{Optimization}\\\textbf{metric}} & \shortstack[c]{\rule{0pt}{1.2em}\textbf{Automatic}\\\textbf{GPU}\\\textbf{allocation}} & \shortstack[c]{\rule{0pt}{1.2em}\textbf{Multi-}\\\textbf{LLM}} & \shortstack[c]{\rule{0pt}{1.2em}\textbf{Unrestricted}\\\textbf{prog.}\\\textbf{model}} & \textbf{Adaptation} & \shortstack[c]{\rule{0pt}{1.2em}\textbf{Profiling}\\\textbf{overhead}} \\
\midrule

\multirow{9}{*}{\rotatebox[origin=c]{90}{\scriptsize\textbf{Multi-LLM serving}}} &
\subgroup{1cm}{4}{Multi-\\plexing} &
AlpaServe~\cite{alpaserve}       & all LLMs & latency SLO & \goodtick & \goodtick & \goodtick & -- & high \\
\lightrule
& & MuxServe~\cite{muxserve}   & all LLMs & throughput & \goodtick & \goodtick & \goodtick & -- & high \\
\lightrule
& & Prism~\cite{prism}         & all LLMs & cost + latency SLO & \goodtick & \goodtick & \goodtick & load & -- \\
\lightrule
& & Aegaeon~\cite{aegaeon}     & all LLMs & latency SLO & \goodtick & \goodtick & \goodtick & load & low \\
\cmidrule{3-10}
&
\subgroup{1cm}{4}{Auto-\\scaling} &
Kubernetes HPA~\cite{kubernetes} & LLM & resource utilization & \goodtick & \goodtick & \goodtick & load & -- \\
\lightrule
& & AIBrix APA~\cite{aibrix}   & LLM & KV cache utilization & \goodtick & \goodtick & \goodtick & load & -- \\
\lightrule
& & llm-d WVA~\cite{llm-d-wva}     & all LLMs & cost + latency SLO & \goodtick & \goodtick & \goodtick & load & low \\
\lightrule
& & vLLM Production Stack~\cite{vllm-prob-stack} & LLM & queue length & \goodtick & \goodtick & \goodtick & load & -- \\

\midrule

\multirow{7}{*}{\rotatebox[origin=c]{90}{\scriptsize\textbf{Agentic Workflow}}} &
\subgroup{1cm}{3}{Static\\analysis} &
Parrot~\cite{parrot}       & workflow request & latency & \badcross  & \badcross  & \badcross & -- & -- \\
\lightrule
& & Ayo~\cite{ayo}             & workflow request & latency & \badcross  & \badcross & \badcross & -- & high \\
\lightrule
& & Murakkab~\cite{murakkab}   & all workflows & energy, cost, accuracy & \goodtick  & \goodtick & \badcross & load & high \\
\cmidrule{3-10}
&
\subgroup{1cm}{3}{Pred.-\\based} &
TokenCake~\cite{tokencake} & workflow request & latency & \badcross  & \badcross  & \badcross & -- & -- \\
\lightrule
& & Agentix~\cite{agentix}   & workflow request & latency & \badcross  & \badcross & \goodtick & -- & -- \\
\lightrule
& & JITServe~\cite{jitserve}      & workflow request & latency SLO & \badcross  & \badcross  & \goodtick & -- & low \\
\cmidrule{3-10}

& & \textbf{\sys{}} & \textbf{all workflows} & \textbf{latency + throughput} & \goodtick & \goodtick & \goodtick & \textbf{load, structure} & \textbf{low} \\
\bottomrule
\end{tabular}
\caption{State-of-the-art systems comparison}
\label{tab:taxonomy}
\end{table*}

\mypar{(3)~Topology-aware fractional GPU placement~(\S\ref{sec:implementation})} It is important to place the multiple heterogeneous LLM replicas and tensor-parallel instances on GPU fractions in a topology-aware fashion, \ie keeping tensor-parallel instances in the same NVLink domain. This is an NP-hard problem, and not supported by existing orchestrators, such as Kubernetes~\cite{kubernetes}.
It is also important to avoid poor fractional allocations, since they prevent future placements due to internal fragmentation.
\sys{} performs topology-aware fractional GPU placement on top of Kubernetes: (1)~it ensures low latency by placing large tensor-parallel LLM fractions first; (2)~it reduces fragmentation by favoring well-balanced tensor-parallel allocations within a GPU node, while packing smaller fractions tightly; (3)~it produces Kubernetes deployment files, which a modified device plugin realizes as hierarchical placements; and (4)~it uses Nvidia MPS to enforce isolation between LLM fractions co-located on a GPU\@.

\tinyskip

\noindent
Our prototype of \sys{}\footnote{Available post-acceptance} is implemented in $25$K lines of Python code, and supports any agentic framework~\cite{langchain, langgraph, camel, autogen} and most LLM execution engines~\cite{vllm, sglang}. We evaluate \sys{} on three agentic workflows (RAG\bplus{}reranker, beam search, and a coding agent) on a $16$-GPU cluster. Compared to the Kubernetes autoscaler~\cite{kubernetes}, \sys{} reduces average workflow latency by $1.0$--$3.3\times$ when running below saturation, and sustains up to $2.5\times$ higher throughput before saturating for the same GPU budget. Compared to workflow-aware serving with Agentix~\cite{agentix}, the latency reduction is $1.0$--$2.0\times$ and the throughput gain up to $2.0\times$. \sys{} also outperforms multi-LLM serving with Aegaeon~\cite{aegaeon} (\Cref{sec:e2e-workflows}).

\section{Agentic Workflow Serving}%
\label{sec:background}

This section states the properties a serving system must handle~(\Cref{subsec:agentic-workflows}), and explains where existing solutions fall short: existing multi-LLM serving systems allocate GPUs across models but do not see the dependencies between them~(\Cref{subsec:multi-llm-serving}), whereas agentic workflow serving systems see the dependencies but schedule requests within an allocation the user fixes~(\Cref{subsec:serving-agentic-workflows}).
We conclude with a driving observation: even though workflow requests are hard to predict, each LLM's share of execution time is stable across requests~(\Cref{subsec:limitations}).

\subsection{Challenges of serving agentic workflows}%
\label{subsec:agentic-workflows}

\mypar{Properties of agentic workflows} A \emph{workflow request} is one end-to-end execution of a workflow for one user input. It issues \emph{LLM invocations}, each a completions request to one LLM, in sequence or in parallel, interleaved with tool calls and orchestration code. Which LLM is invoked next, and how often, depends on earlier outputs. Two properties of agentic workflows matter for serving: (1)~they are \emph{multi-LLM}, combining LLMs of different sizes in different roles~\cite{frugalgpt, openrouter, copilot-auto, superpowers-sdd, test-time-scaling, beam-with-prm, rag-reranker, flashrag}; and (2)~their execution is \emph{unpredictable}, because execution paths and times depend on the request input. In our beam search trace, end-to-end request latency spans $9$--$264$\unit{s}, and generator invocations per request $24$--$844$.

\mypar{Diversity of agentic programming models} There are multiple frameworks to write agentic workflows, each with its own API and programming model~\cite{agent-programming-models, langgraph, llamaindex, autogen, camel}, but across all frameworks, each LLM invocation reaches a serving engine as a completions request. A serving system can therefore learn a workflow's structure from information provided by a framework, requiring programs to be written against that specific framework, or from the stream of requests, which requires no change to the program.

Workflow properties and framework diversity translate into six requirements against which we compare existing systems~(\Cref{tab:taxonomy}). A serving system must: (1)~optimize the \emph{target and metric} users see, the end-to-end latency and throughput of all workflows, rather than of one LLM, all LLMs jointly, or a single workflow request; (2)~\emph{allocate GPUs} to LLMs automatically, because sizing an allocation across LLMs of different sizes by hand wastes GPUs; (3)~support \emph{multiple LLMs}; (4)~accept workflows written against any framework, without prescribing a \emph{programming model}; (5)~\emph{adapt} the allocation when the load (arrival rate) or the workflow structure (LLM invocations per request) shifts, because a fixed allocation will over-provision some LLMs and under-provision others; and (6)~need limited \emph{profiling} before it can serve a workflow, without re-profiling across workload shifts, because profiling delays serving and adaptation on every LLM\@.

\subsection{Approaches for multi-LLM serving}%
\label{subsec:multi-llm-serving}

Multi-LLM serving systems run inference for several LLMs on one cluster. They either \emph{autoscale} or \emph{multiplex} models across the cluster GPUs, but without knowledge of the dependencies between the models. Ignoring the workflow makes them general: they impose no programming model, accept any mix of LLMs, and allocate GPUs without user input~(\Cref{tab:taxonomy}).

\myparr{Autoscaling systems} adjust the number of replicas of each LLM on demand. They scale either on a resource metric of that LLM's replicas~\cite{kubernetes, vllm-prob-stack, aibrix}, such as GPU utilization, KV cache usage or queue length, or on a policy that targets a service-level objective~(SLO) on latency~\cite{aibrix, llm-d-wva}. An autoscaler can only control replica count, and therefore cannot trade throughput for latency via tensor parallelism. The autoscaling signal also lags behind the workload: a full queue triggers a scale-up, the queue drains, and a later reading triggers a scale-down, resulting in unstable allocations~(\Cref{sec:e2e-workflows}).

\myparr{Multiplexing systems} co-locate multiple LLMs on the same set of GPUs, either through temporal multiplexing~\cite{alpaserve}, spatial multiplexing~\cite{prism} or a mix of both~\cite{muxserve, aegaeon}. Their objectives are model-level: SLO attainment~\cite{alpaserve}, aggregate throughput~\cite{muxserve}, or the fewest GPUs that meet per-model SLOs~\cite{prism, aegaeon}. None of them sees workflow-level dependencies between LLM invocations, so a user must manually convert a workflow-level target into per-model SLOs.

\subsection{Approaches for agentic workflow serving}%
\label{subsec:serving-agentic-workflows}

Agentic workflow serving systems exploit the workflow-level dependencies between LLM invocations, but they are limited to scheduling requests within a given GPU allocation and assume a single LLM, with Murakkab~\cite{murakkab} as the exception on both counts~(\Cref{tab:taxonomy}). These systems can be classified into two categories: \emph{static analysis} approaches and \emph{prediction-based scheduling} approaches.

\myparr{Static analysis systems} propose custom abstractions for agentic programming, in order to expose static dataflow graphs of agentic workflows. Analyzing workflow graphs enables them to apply new optimizations. For example, Parrot~\cite{parrot} applies graph-aware request batching and KV cache sharing across requests to reduce inference latency. Ayo~\cite{ayo} iteratively transforms the workflow graph with optimizations, such as splitting the prefill phase for partial prompts and pipelining decoding of semantically divisible tasks. A declarative specification can go further. Because it separates what a workflow does from how it runs, Murakkab~\cite{murakkab} selects the model, the hardware, and the parallelism of each step, together with the number of instances to deploy. It optimizes energy, cost, or accuracy rather than throughput and latency. The price is the abstraction: the workflow must be written against it, which rules out applications already written against an existing agentic framework~(\Cref{subsec:agentic-workflows}).

\myparr{Prediction-based scheduling systems} formalize agentic workflows as dynamic graphs, and use workflow-level statistics to forecast request completion time and make request scheduling decisions. For example, Agentix~\cite{agentix} uses cumulative LLM execution time to assign request priorities, and dynamically preempts low-priority requests to avoid head-of-line blocking and improve overall latency. Similarly, JITServe~\cite{jitserve} estimates an upper bound on response length, and gradually refines it to schedule serving bandwidth with respect to latency SLOs. TokenCake~\cite{tokencake} predicts when an agent returns from a function call and moves its KV cache off the GPU meanwhile. All three schedule requests within a fixed allocation, so a bottleneck LLM stays the bottleneck whatever the order.

\begin{figure}[tb]
  \centering\small
  \begin{subfigure}[b]{0.42\linewidth}
    \centering
    \includegraphics[width=\linewidth]{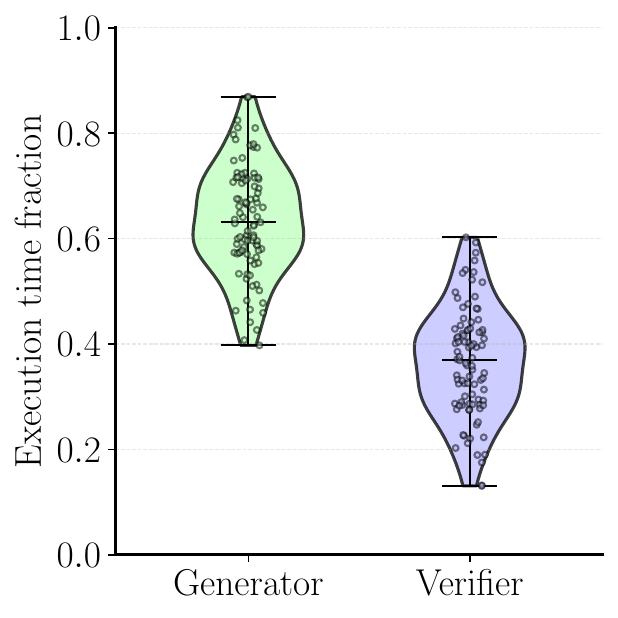}
    \caption{Beam search}
    \label{fig:fraction-execution-time}
  \end{subfigure}\hfill
  ~
  \begin{subfigure}[b]{0.55\linewidth}
    \centering
    \includegraphics[width=\linewidth]{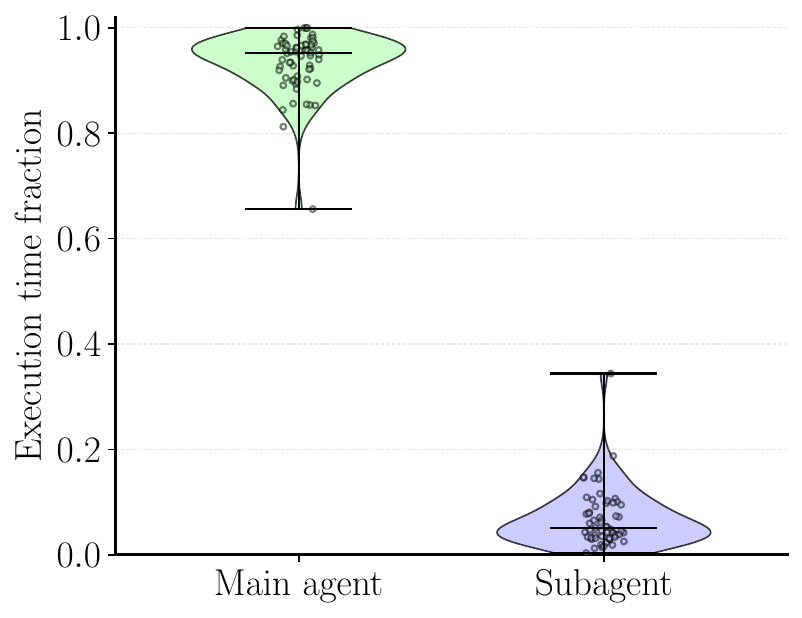}
    \caption{Coding agent}
    \label{fig:fraction-execution-time-coding-agent}
  \end{subfigure}
  \caption{Share of each LLM's execution time, across requests}
\end{figure}

\subsection{Stable per-LLM execution-time shares}
\label{subsec:limitations}

To allocate GPUs across the LLMs of a workflow, a serving system must know how much of the workflow's demand each LLM accounts for. Per-request measurements cannot tell it: across the requests of our beam search trace, the number of generator invocations and the end-to-end latency each vary about $30\times$~(\Cref{subsec:agentic-workflows}). No single request is representative, and an allocation must hold across many.

The share of a request's LLM execution time that each LLM accounts for, however, is stable: over $91$~requests, the generator's share stays between $40$\% and $88$\%, with a median of $64$\%~(\Cref{fig:fraction-execution-time}). The reason is structural: the LLMs of a workflow are coupled. In beam search, every generator invocation is followed by a verifier invocation, so a request that is slow in one is slow in the other. The coding agent shows the same pattern. Its main agent issues a sequence of calls over a context that grows with the session and spawns sub-agents on a smaller LLM. A single main-agent invocation takes between $1$\unit{s} and $91$\unit{s}. Across $62$~requests, however, the main agent's share of LLM execution time has a median of $95$\% and stays above $80$\% in all but one request~(\Cref{fig:fraction-execution-time-coding-agent}). The relative statistic thus singles out the main agent as the bottleneck LLM.

\section{\sys{} Overview}
\label{sec:overview}

\begin{figure*}[t]
    \centering
    \includegraphics[width=0.8\textwidth]{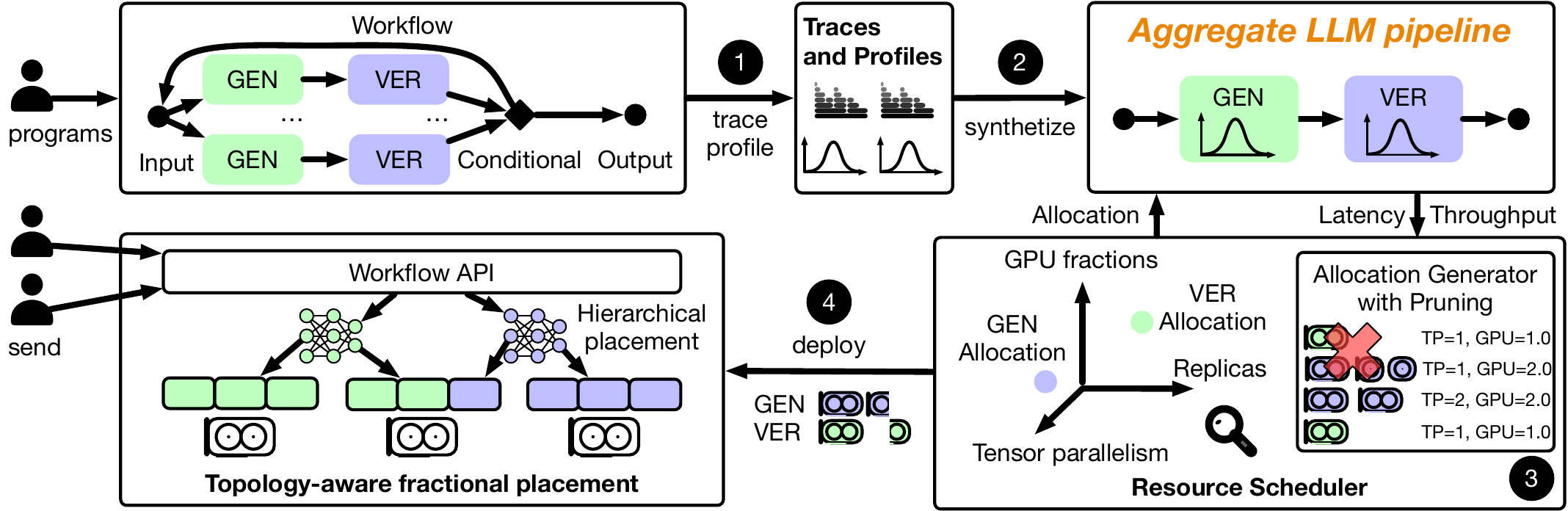}
    \caption{\sys{} Overview}
    \label{fig:overview}
\end{figure*}

To allocate GPUs across a workflow's LLMs, a serving system must predict how an allocation changes workflow latency and throughput. No single request supports the prediction: its control flow, LLM invocation count, and sequence lengths depend on the input~(\Cref{subsec:agentic-workflows}). \sys{} builds on the observation that each LLM's share of a request's LLM execution time is stable across requests, although end-to-end latency varies about $30\times$ in beam search~(\Cref{subsec:limitations}). It therefore sizes the allocation to each LLM's average demand per request, not to any one execution path.

\sys{} captures this average demand in an \emph{\alp{}}, a pipeline with one stage per LLM. A stage combines the LLM's workflow statistics, its average invocations per workflow request and average invocations in flight at once, with a throughput--latency profile of the LLM measured in isolation. For an allocation, the GPU fraction, tensor parallelism degree, and replica count of every LLM, the \alp{} predicts workflow latency at a target arrival rate and maximum throughput by profile lookups and arithmetic. The GPU scheduler thus compares allocations without profiling any.

\Cref{fig:overview} shows how \sys{} serves one or more workflows on a GPU cluster in four steps: \myc{1}~it traces the LLM invocations of each workflow and profiles each LLM by replaying them; \myc{2}~it synthesizes traces and profiles into one \alp{} per workflow; \myc{3}~a GPU scheduler searches for the allocation with the lowest predicted latency at every workflow's target arrival rate; and \myc{4}~it places the chosen replicas, keeping each tensor-parallel replica within one NVLink domain and enforcing each GPU fraction.

Static-analysis serving systems learn a workflow's structure from its program, so the workflow must be written against their abstraction~(\Cref{subsec:serving-agentic-workflows}). \sys{} learns it from the stream of LLM invocations instead, so any agentic framework works~(\myc{1}). An HTTP proxy in front of each serving engine's completions API records every invocation's request and response contents, start and end timestamps, and workflow request. From the traces, \sys{} extracts each LLM's invocations per workflow request and, from overlapping timestamps, its invocations in flight at once. It then profiles each LLM by replaying its traced invocations, sweeping the arrival rate to saturation at every tensor parallelism degree the scheduler may choose. Replicas need no sweep: throughput scales linearly with them and latency is unaffected.

\sys{} synthesizes workflow statistics and profiles into the \alp{}~(\myc{2}). A workflow arrival rate induces an LLM arrival rate of that rate times the LLM's invocations per request, divided among its replicas. The LLM's latency contribution is its profiled latency at that rate, times the invocations per request, divided by the invocations in flight, since those overlap. Loops and fan-out thus fold into one stage per LLM: in \Cref{fig:overview}, the looped and parallel invocations of the generator and the verifier become the \textsf{GEN} and \textsf{VER} stages. Workflow latency is the sum of the stage contributions, so stage order is irrelevant, and maximum throughput is the smallest per-LLM throughput divided by that LLM's invocations per request~(\Cref{sec:performance-model}).

Speeding up the bottleneck LLM moves the bottleneck to the next stage, so LLMs cannot be sized one at a time. The GPU scheduler therefore searches jointly over all LLMs of all deployed workflows~(\myc{3}). Per LLM it chooses a GPU fraction (so that small LLMs share a GPU), a tensor parallelism degree, and a replica count, and picks the allocation with the \emph{lowest predicted latency that sustains every workflow's target arrival rate}. Three rules prune the search: a higher-latency LLM never gets fewer fractions than a lower-latency one, fractions are packed contiguously onto GPUs, and tensor parallelism is capped at the NVLink domain size. With $16$~GPUs and up to $5$~LLMs, the search takes under $35$\unit{s}~(\Cref{sec:scheduler-latency}).

\sys{} places the chosen replicas itself~(\myc{4}), because a tensor-parallel replica must sit within one NVLink domain, a fractional replica needs a hard limit on its GPU share, and no existing orchestrator supports both~(\Cref{sec:implementation}). A two-stage heuristic assigns replicas to nodes, largest tensor-parallel replicas first so that small fractions do not fragment NVLink domains, then GPU fractions within each node. Kubernetes deployment files pin the node, an extended Nvidia device plugin pins the fractions, and Nvidia MPS enforces each fraction's SM and memory limits.

A deployed workflow's workload shifts: the arrival rate changes, a parameter such as the beam width changes the invocations per request, and workflows are deployed or retired. \sys{} therefore closes a feedback loop from serving to the \alp{} and the scheduler. Live traces refresh the workflow statistics of the \alp{}, while the per-LLM profiles stay valid, so no LLM is profiled again. When the observed arrival rate or workflow statistics deviate from those the allocation was computed for, beyond a threshold and over several consecutive windows, the scheduler reruns its search on the refreshed \alp{}. It reconfigures the cluster only when the predicted gain exceeds a threshold, restarting only the replicas whose allocation changed~(\Cref{sec:scheduling:adaptation}).

\begin{figure*}[t]
    \centering
    \includegraphics[width=\textwidth]{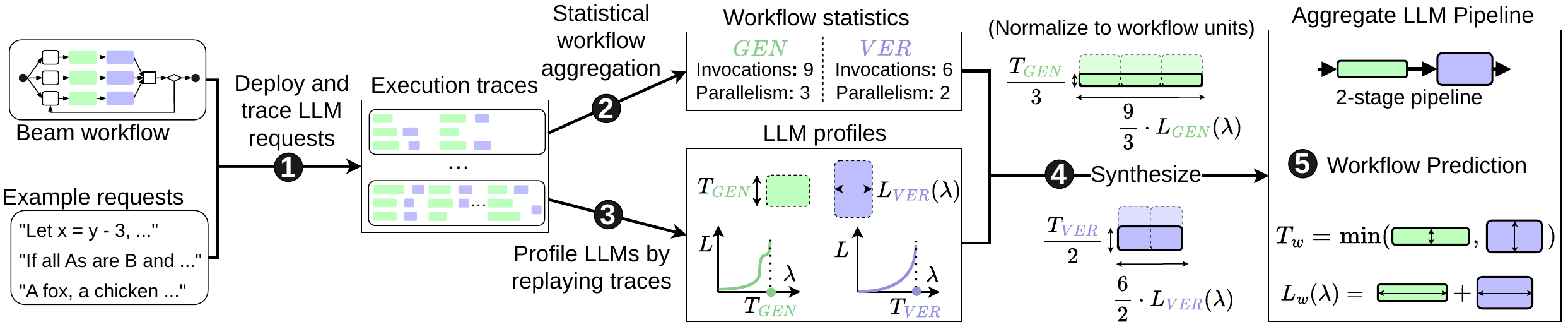}
    \caption{Construction of an \alp{} for a beam search workflow}
    \label{fig:alp-diagram}
\end{figure*}

\section{Aggregate LLM Pipeline}
\label{sec:performance-model}

An \alp{} abstracts a workflow into a pipeline with one stage per LLM. Each stage carries per-LLM statistics that are stable across workflow requests, although a single request is not~(\Cref{subsec:limitations}), which is what lets \sys{} allocate GPUs to a workflow whose control flow it cannot predict. The stages come from execution traces, not source code, so any agentic framework and any engine with a completions API works. One stage per LLM admits heterogeneous LLMs. Each stage is a function of the GPUs its LLM receives, so \sys{} scores an allocation before deploying it, and each stage carries a latency and a throughput, so the GPU scheduler~(\Cref{sec:resource-scheduling}) can trade one for the other. \Cref{fig:alp-diagram} shows the five steps that build an \alp{} and evaluate it for a candidate allocation.

\mypar{\myc{1}~Workflow tracing} \sys{} deploys the workflow on any allocation and runs a sample of workflow requests through it. The result is one \emph{execution trace} per workflow request: its LLM invocations, each with prompt, completion, start and end timestamps, and target LLM. An HTTP proxy in front of the completions API of each serving engine records them. Every framework reaches the engine through this API~(\Cref{subsec:agentic-workflows}), so neither the framework nor the engine changes. The proxy tags each invocation with the workflow request that issued it. The timestamps serve only to tell which invocations of one workflow request overlap, not how fast this deployment ran, so \sys{} can trace workflow requests in parallel, each in its own deployment.

\mypar{\myc{2}~Workflow aggregation} For every LLM~$m$, \sys{} extracts two \emph{workflow statistics}, averaged over the traced workflow requests: the invocations per workflow request, $n_m$, and the request-level parallelism, $p_m$, the number of invocations of $m$ in flight at once within one workflow request, averaged over the time during which at least one runs. In \Cref{fig:alp-diagram}, the generator expands three beams at a time over nine invocations, so $n_{\textsf{GEN}} = 9$ and $p_{\textsf{GEN}} = 3$. The verifier scores two beams at a time over six invocations, because the generator produces the same value on two of the three beams, so $n_{\textsf{VER}} = 6$ and $p_{\textsf{VER}} = 2$. A loop raises $n_m$, a fan-out raises $p_m$, and a branch that skips $m$ lowers $n_m$, without any analysis of the program. Tool calls and database lookups appear only as gaps between invocations. \sys{} drops them: they consume no GPU, and their time is the same under every allocation, so it does not change the ranking. Tools whose run time dominates a request are not modeled. The statistics also fix the relation between workflow-level and LLM-level load: a workflow arrival rate $\lambda_w$ induces an arrival rate $\lambda_m = \lambda_w \cdot n_m$ at LLM~$m$.

\mypar{\myc{3}~LLM profiling} \sys{} measures a \emph{throughput--latency profile} per LLM and tensor parallelism degree, which predicts how the LLM performs on the GPUs an allocation gives it. It replays the invocations of $m$ from all traces against a deployment of $m$ alone, at rising arrival rates in invocations per second. Within a trace, an invocation is issued only after those it followed have completed, so the replay reproduces the in-flight count $p_m$, and each invocation is forced to the recorded completion length. Replaying traced invocations therefore reproduces the prompt and completion lengths of the workload, hence its ratio of prefill to decode, and its prefix sharing, which gives beam search a high prefix cache hit rate because every round extends the previous prompt. The sweep runs from low load to saturation, the point of maximum throughput, and records that throughput as $T_m$. At each rate, \sys{} records the average latency and the 50th, 90th, and 99th percentiles. The sweep is repeated at every tensor parallelism degree the scheduler may choose~(\Cref{sec:resource-scheduling}), because latency and throughput do not scale linearly with the degree, but not over replicas: assuming the router spreads invocations evenly, replicas split the arrival rate and their saturation throughputs add up. Profiles of different LLMs are independent, so \sys{} measures them concurrently on disjoint GPUs.

\mypar{\myc{4}~\alp{} synthesis} \sys{} synthesizes the workflow statistics and per-LLM profiles into a pipeline of unique LLMs that captures the agentic workflow's throughput--latency characteristics. \sys{} does so by transforming each LLM's throughput--latency profile from LLM-level to workflow-level units. In the resulting pipeline, the latency contribution of each LLM~$m$ to the latency of workflow~$w$ incorporates both the number of invocations and request-level parallelism statistics of model $m$:
\begin{equation}\label{eq:latency}
    L_w(\lambda_w) = \sum_{m} L_{w_m}(\lambda_w) = \sum_{m} L_m(\lambda_w \cdot n_m) \cdot \frac{n_m}{p_m}
\end{equation}
where $\lambda_w$ is the target workflow-level arrival rate, and $L_m$ is the average per-request latency of LLM~$m$ at the LLM arrival rate~$\lambda_w \cdot n_m$. The ratio~$n_m / p_m$ scales the LLM's latency to account for its parallelism. The workflow latency~$L_w$ is therefore the sum over all LLMs, while the maximum workflow throughput~$T_w$ is the minimum across LLMs of the throughput that each can sustain, normalized to workflow requests:
\begin{equation}\label{eq:throughput}
    T_w = \min_{m} \frac{T_m}{n_m}
\end{equation}
where $T_m$ is the maximum throughput of LLM~$m$. Request-level parallelism ($p_m$) and arrival rate ($\lambda_w$) are not used here, since parallel requests do not reduce the total number of requests for an LLM, and the maximum throughput is independent of the arrival rate (respectively). The resulting throughput--latency curve of LLM $m$ is expressed in terms of workflow-level throughput ($\lambda_w$) and workflow-level latency contribution ($L_{w_m}$), which is also parameterized by the degree of tensor parallelism and target latency percentile when selecting the LLM profile data (not shown for clarity). Stage order enters neither equation, since a sum and a minimum are invariant under permutation. An \alp{} therefore does not model the order in which a workflow invokes its LLMs, which is what lets it hold for data-dependent control flow. Once all LLMs are expressed in workflow-level units, each stage carries two numbers and the scheduler reads both: the LLM with the largest $L_{w_m}$ dominates workflow latency, the LLM with the smallest $T_m / n_m$ caps workflow throughput, and the ratios of the $L_{w_m}$ order the LLMs by GPU demand~(\Cref{sec:resource-scheduling}).

\mypar{\myc{5}~Workflow prediction} A candidate allocation gives each LLM~$m$ a tensor parallelism degree~$TP_m$ and a replica count~$d_m$. With the target arrival rate~$\lambda_w$ and the percentile at which latency is to be predicted, this is everything the prediction needs. \sys{} selects each LLM's profile at $TP_m$, on the average-latency or the requested percentile curve, as $L_m$ and $T_m$, scales the arrival rate to $\lambda_w \cdot n_m / d_m$ for $d_m$~replicas, and combines the per-LLM values with \Cref{eq:latency,eq:throughput}. An allocation is feasible only if $\lambda_w \le T_w$, that is, if no stage saturates at the target rate. A prediction thus costs one profile lookup and a few arithmetic operations per LLM, against a deployment and a profiling run per candidate, which lets the GPU scheduler score every allocation its search enumerates~(\Cref{sec:scheduler-latency}). Both equations predict steady state from averaged statistics. The averages do not hold for fan-out across different LLMs, which \Cref{eq:latency} treats as sequential, or for an LLM used in two roles with distinct load.

\section{GPU Scheduling}\label{sec:resource-scheduling}

For each agentic workflow, the GPU scheduler must allocate the right amount of GPU resources to each LLM. The LLMs are coupled through the workflow, so they cannot be sized one at a time, and the space of joint allocations grows with the number of LLMs, of GPUs, and of parallelism configurations per LLM. \Cref{sec:scheduling:search} describes how the scheduler searches this space for an allocation at a target arrival rate without enumerating all of it, and \Cref{sec:scheduling:adaptation} how \sys{} detects workload shifts and reruns the search.

\subsection{Searching for an allocation}\label{sec:scheduling:search}

The GPU scheduler ranks candidate allocations by a \emph{utility} that the \alp{} predicts~(\Cref{sec:performance-model}). The utility is $-\infty$ if the predicted workflow throughput~$T_w$ falls short of the target arrival rate~$\lambda_w$, since an LLM then saturates, and the negated predicted workflow latency~$-L_w(\lambda_w)$ otherwise. Throughput is thus a constraint on the search and latency its objective. An allocation fixes three quantities per LLM: a replica count, a tensor parallelism degree, and a GPU share.

The LLMs of a workflow differ in size, such as the embedding LLM and the generative LLM of RAG\bplus{}reranker, and so in GPU demand. The scheduler therefore allocates \emph{GPU fractions}, slices of a single GPU, so that small LLMs share a GPU. The \emph{GPU share} of an LLM is the sum of its fractions and can exceed one GPU. With an embedding LLM of latency $1$\unit{s} and a generative LLM of $12$\unit{s}, whole-GPU granularity on $4$~GPUs gives the embedding LLM $1/4$ of the resources instead of its $1/13$ share.

\begin{figure}[t]
    \centering
    \includegraphics[width=0.9\columnwidth]{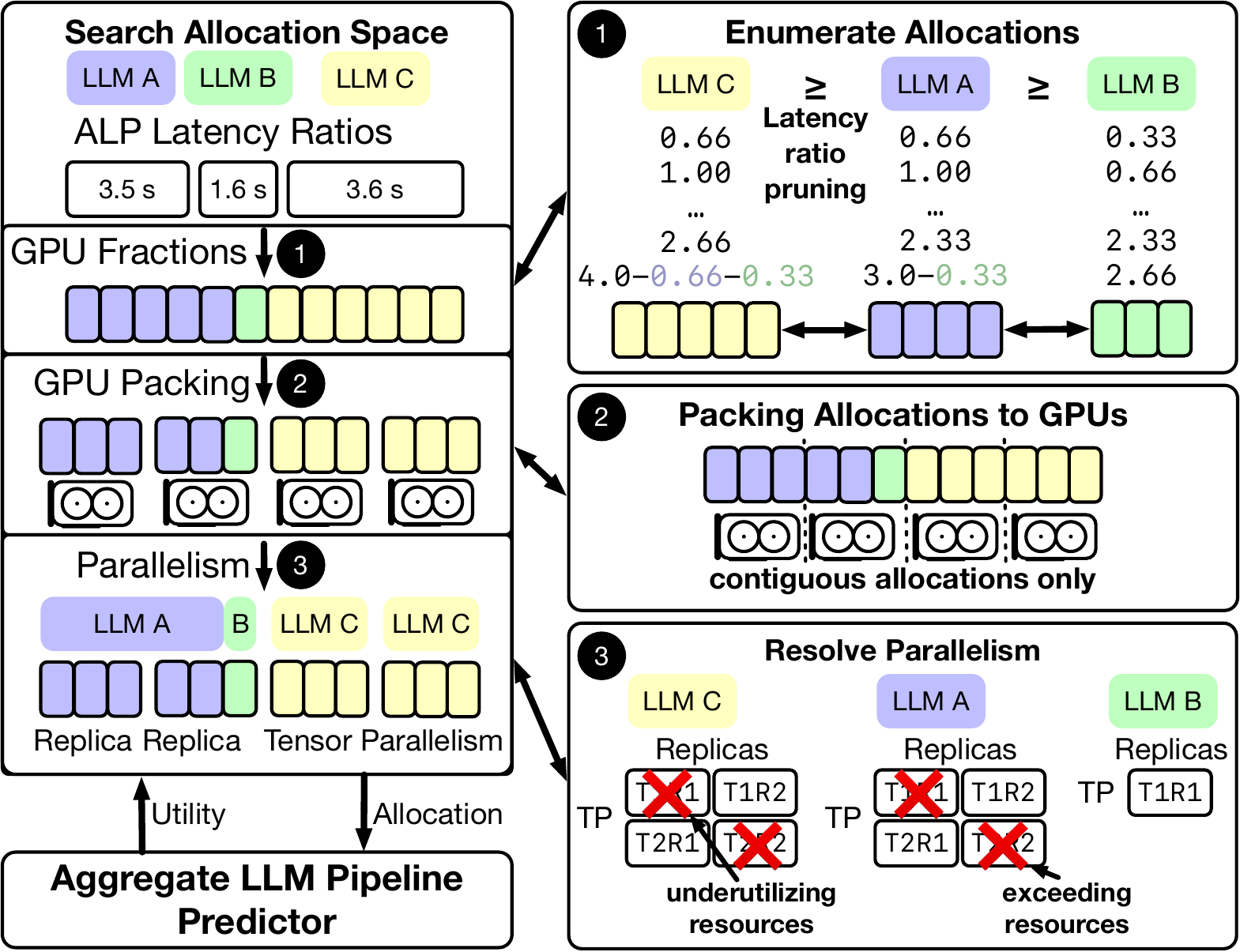}
    \caption{Overview of GPU scheduling in \sys}
    \label{fig:scheduling-algorithm}
\end{figure}

\Cref{fig:scheduling-algorithm} shows the search. From the latency ratios of the LLMs in the \alp{}, the scheduler \myc{1}~enumerates a GPU share per LLM, \myc{2}~packs the shares onto GPUs, and \myc{3}~enumerates the parallelism configurations that fit each LLM's GPUs. Each step prunes the space. The \alp{} scores every surviving allocation, and \sys{} deploys the one with the highest utility through topology-aware fractional placement~(\Cref{sec:implementation}). Pruning keeps the search under $35$\unit{s} for $16$~GPUs and up to $5$~LLMs~(\Cref{sec:scheduler-latency}).

\mypar{\myc{1}~Enumeration of GPU shares} The first step enumerates assignments of GPU fractions to LLMs. With $5$~LLMs, $16$~GPUs, and $10$~fractions per GPU, there are $\binom{16 \cdot 10 + 5 - 1}{5 - 1} \approx 29$~million before any parallelism configuration is considered, so two rules prune the step. First, the scheduler assigns shares in latency order, so that an LLM with a higher latency ratio never gets a smaller share. Latency is a proxy for GPU demand, so the order excludes giving an embedding LLM more resources than a generative LLM, while the boundary between two shares stays free to move. Second, each share is bounded below by the fractions the LLM needs to load its parameters and initialize its KV cache, and above by the fractions not yet assigned to higher-latency LLMs, less the lower bounds of all lower-latency LLMs.

\mypar{\myc{2}~Packing shares onto GPUs} The second step maps the shares to GPUs, because the third step needs the number of GPUs each LLM spans. It packs the shares contiguously in latency order, so that it considers one packing per share assignment rather than every order of LLMs and GPUs. An LLM with a share of $1.66$~GPUs takes all of one GPU and $0.66$ of the next. Which physical GPUs it holds is left to placement~(\Cref{sec:implementation}). The latency order hands the highest-latency LLMs whole GPUs before any GPU is split, so they are the least partitioned across GPUs.

\mypar{\myc{3}~Resolving parallelism} Given the number of GPUs each LLM spans, the third step keeps the combinations of tensor parallelism degree and replica count whose product equals the span. A larger product does not fit, and a smaller one leaves GPUs unused. It also drops degrees above the number of GPUs in an NVLink domain, because the replica would then communicate over a slower interconnect at every layer.

\mypar{Scoring a candidate allocation} The three steps fix a GPU share~$f_m$, a tensor parallelism degree~$TP_m$, and a replica count~$d_m$ per LLM~$m$. The degree and the replica count enter the latency prediction as before~(\Cref{sec:performance-model}). The share enters the throughput. \sys{} profiles one replica of LLM~$m$ on $TP_m$ whole GPUs, so a share of $f_m$ is worth $f_m / TP_m$ profiled replicas, and \Cref{eq:throughput} becomes
\begin{equation}\label{eq:throughput-fractional}
    T_w = \min_{m} \frac{f_m}{n_m \cdot TP_m} \cdot T_m
\end{equation}
The replica count drops out, since only the total share decides how much throughput an LLM sustains, and its replicas need not hold equal shares. The latency profile stays unscaled: a smaller share lowers how many requests a replica absorbs before it saturates, not how fast it serves them below saturation, and the throughput constraint keeps every replica below saturation. Scoring thus stays one profile lookup per LLM.

\mypar{Scheduling multiple workflows} When a cluster serves several workflows at once, the scheduler adds a level to the search. It first decides how many GPUs each workflow receives, then searches each workflow's allocation within those GPUs as above. It aggregates the per-workflow utilities under an egalitarian welfare policy, which maximizes the utility of the worst-served workflow. Other policies, such as social welfare, which maximizes the sum, fit the same search.

\subsection{Adapting to workload shifts}\label{sec:scheduling:adaptation}
An allocation is computed for one workload: the workflow arrival rate~$\lambda_w$ and the workflow statistics~$n_m$ and~$p_m$ of each LLM. None of them stays fixed: arrival rates change, a parameter such as the beam width changes the structure, and workflows are deployed and retired. \sys{} treats a shift as the scheduling problem it has already solved once, with refreshed inputs. The \alp{} separates what shifts from what does not. The per-LLM profiles~(\Cref{sec:performance-model}) depend on the LLM, the hardware, and the requests, not on how often the workflow invokes the LLM, so they survive a shift in rate or structure, though not one in the requests themselves. The workflow statistics are what shifts, and the tracing proxy records live requests too, so refreshing the \alp{} costs a pass over recent traces, not a profiling run.

The \alp{} also predicts the latency and throughput of any allocation under any workload, so \sys{} knows, before it touches the cluster, whether and by how much a different allocation would serve the shifted workload better. A load signal only says that something changed. The predicted gain says whether the change is worth a reconfiguration. A controller runs alongside serving. It estimates the arrival rate and the workflow statistics from the live traces over a fixed window and compares them with the workload the allocation was computed for. A shift is a deviation that exceeds a threshold over several consecutive windows, which ignores short bursts and catches slow drifts.

On a shift, the controller refreshes the \alp{} and reruns the joint search over all deployed workflows~(\Cref{sec:scheduling:search}), so a shift in one workflow can change the allocation of another. It applies the new allocation only when its predicted gain over the current one exceeds a threshold, because a reconfiguration restarts replicas, and it restarts only the replicas whose allocation changed~(\Cref{sec:implementation}). It then lets the workload settle before it takes the observed workload as the new reference. In the evaluation, a new allocation is in place $138$--$369$\unit{s} after a shift in rate or structure, once per shift~(\Cref{sec:eval:adaptation}).

\section{Topology-Aware Fractional Placement}\label{sec:implementation}

\begin{figure*}[t]
    \centering
    \setlength{\tabcolsep}{1pt}
    \begin{tabular}{@{}c m{0.26\textwidth} m{0.27\textwidth} m{0.26\textwidth}@{}}
        & \multicolumn{1}{c}{\footnotesize\textbf{4 GPUs}}
          & \multicolumn{1}{c}{\footnotesize\textbf{8 GPUs}}
          & \multicolumn{1}{c}{\footnotesize\textbf{16 GPUs}} \\[2pt]

        \rotatebox[origin=c]{90}{\footnotesize\textbf{RAG + Reranker}} &
        \centering\gridpanel{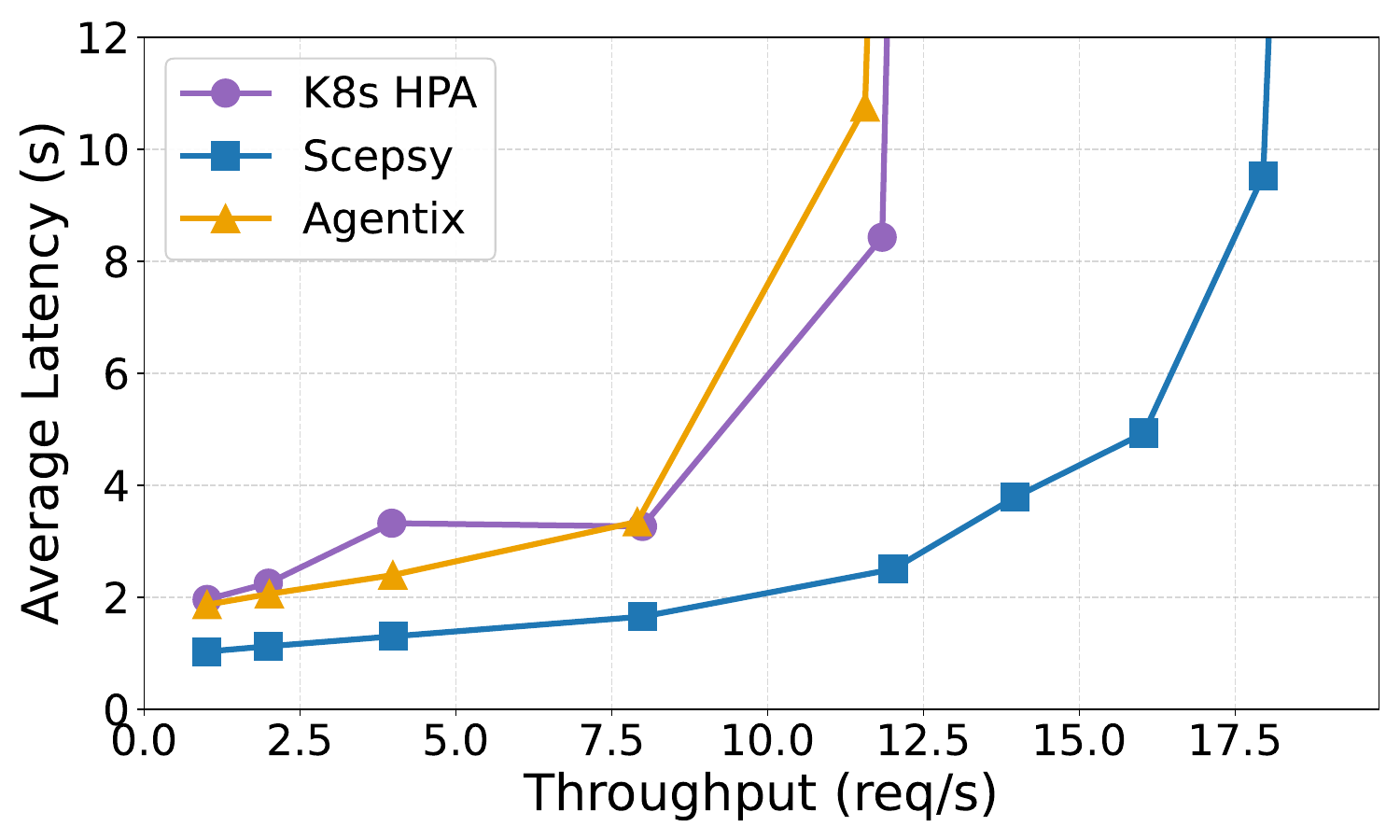}\label{fig:tl-rag-4gpu} &
        \centering\gridpanel{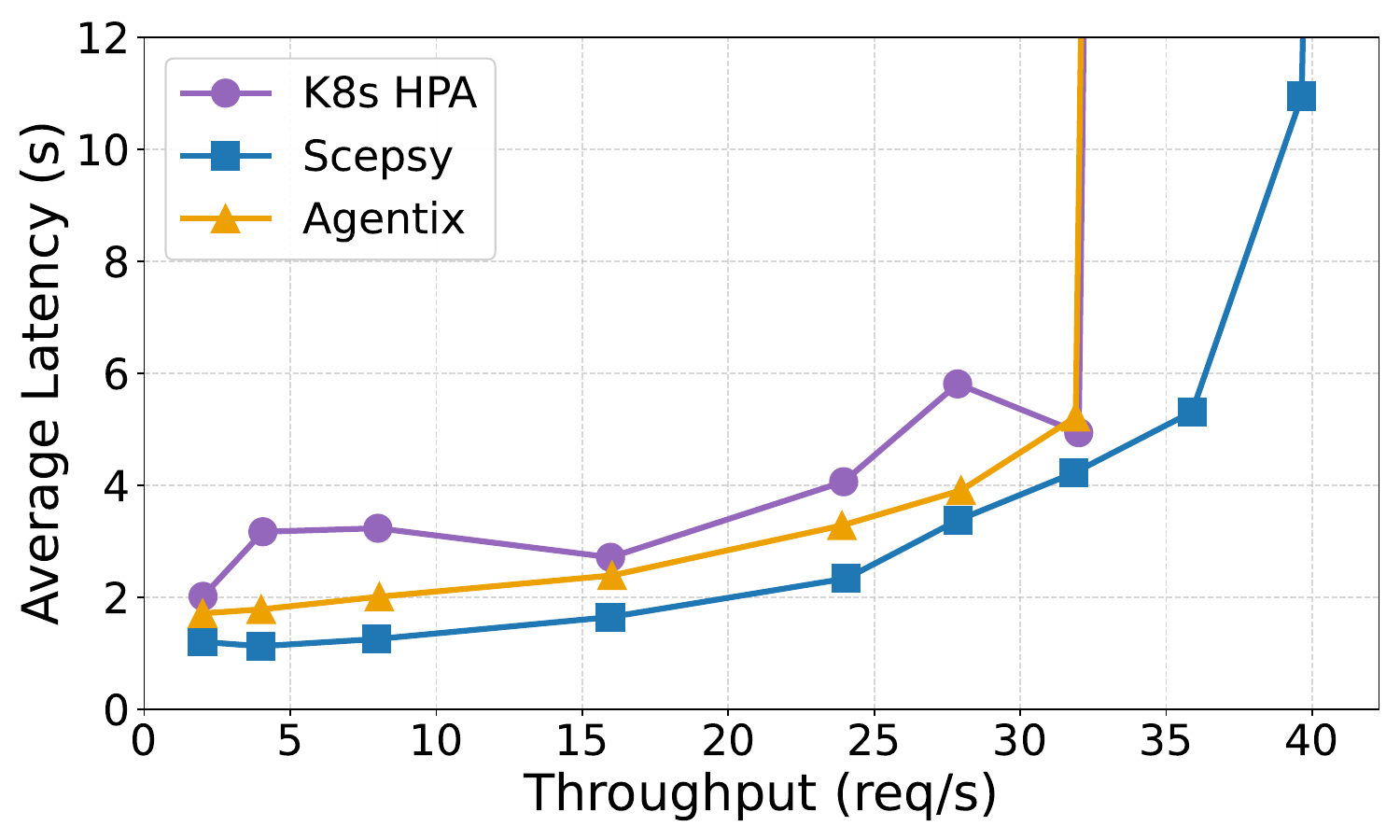}\label{fig:tl-rag-8gpu} &
        \centering\gridpanel{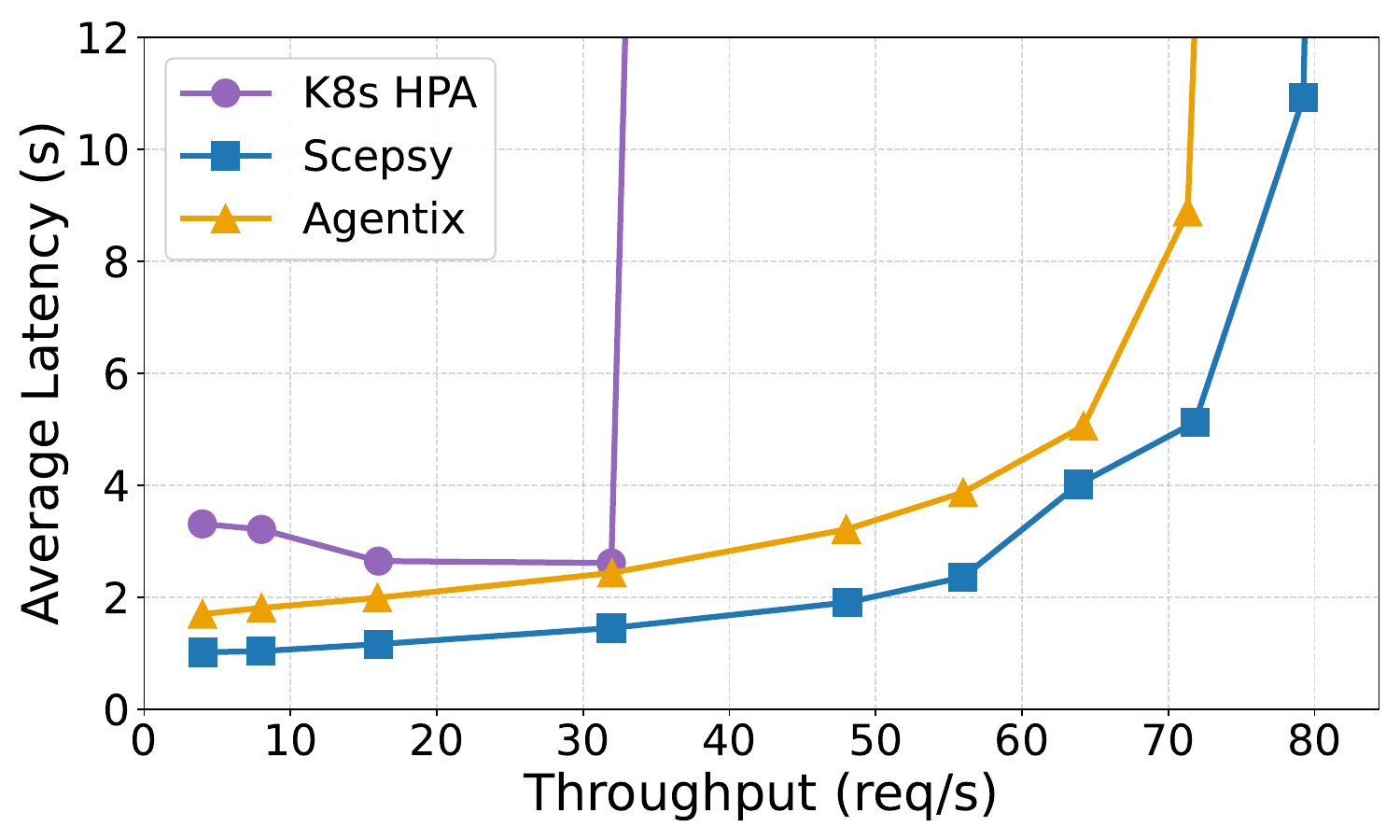}\label{fig:tl-rag-16gpu}
        \tabularnewline

        \rotatebox[origin=c]{90}{\footnotesize\textbf{Beam Search}} &
        \centering\gridpanel{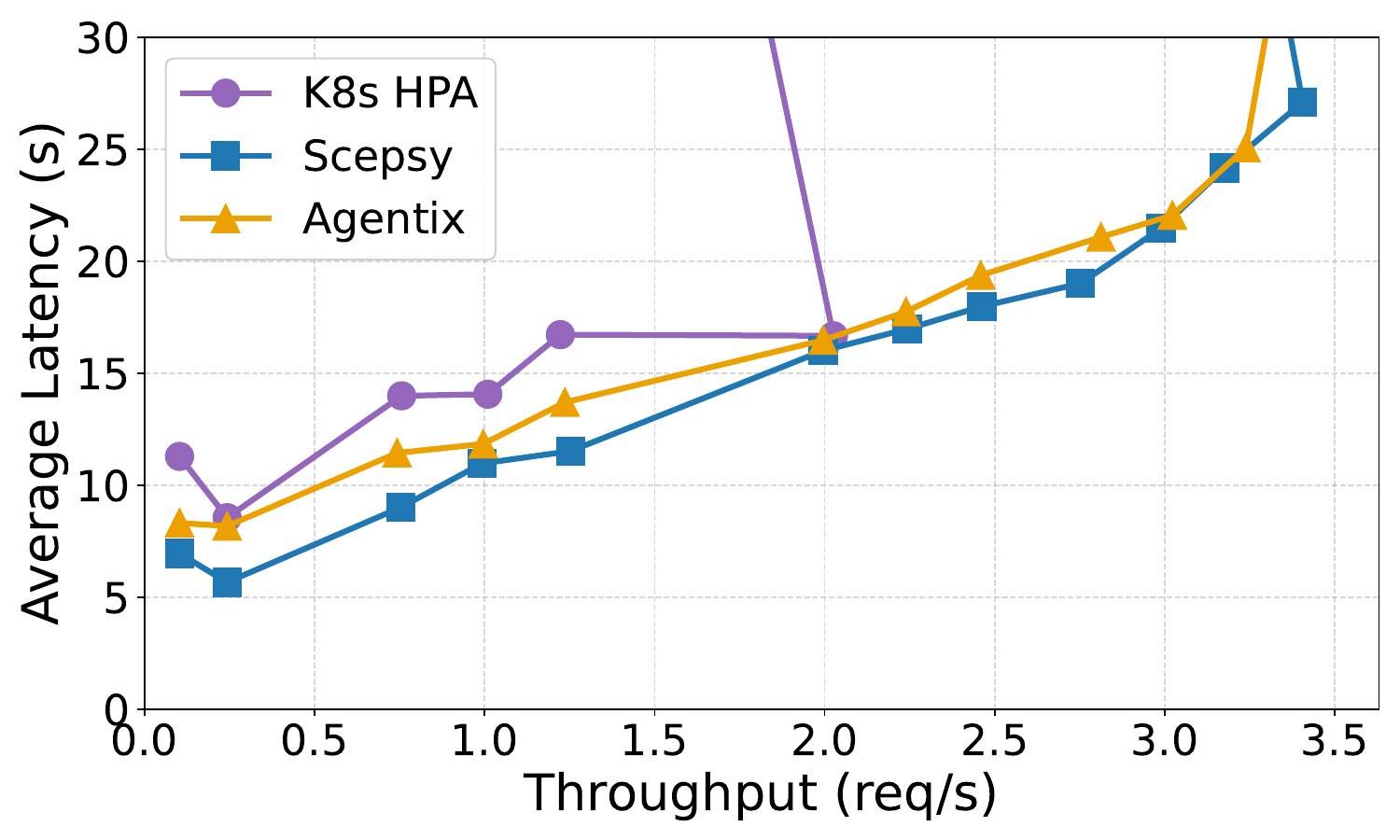}\label{fig:tl-beam-4gpu} &
        \centering\gridpanel{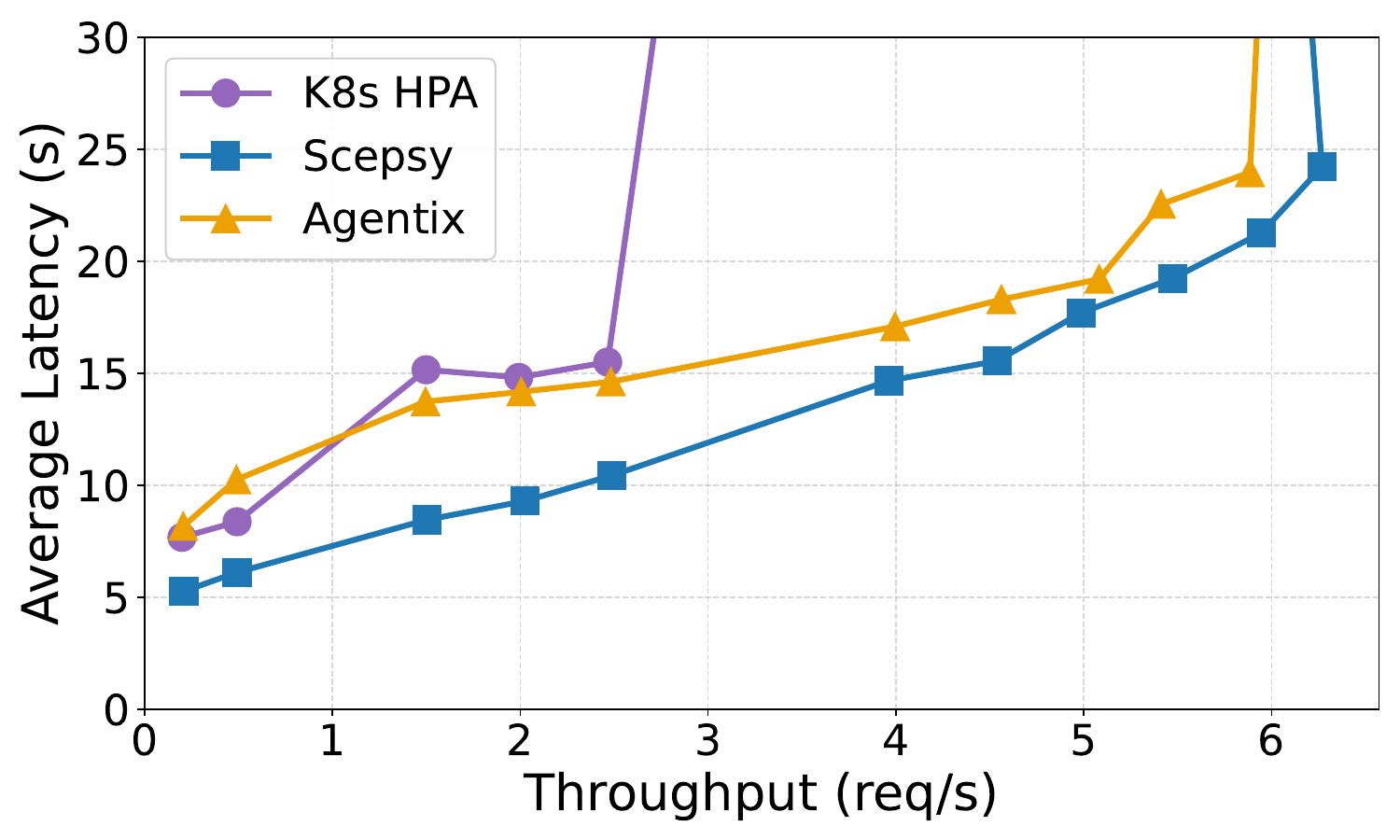}\label{fig:tl-beam-8gpu} &
        \centering\gridpanel{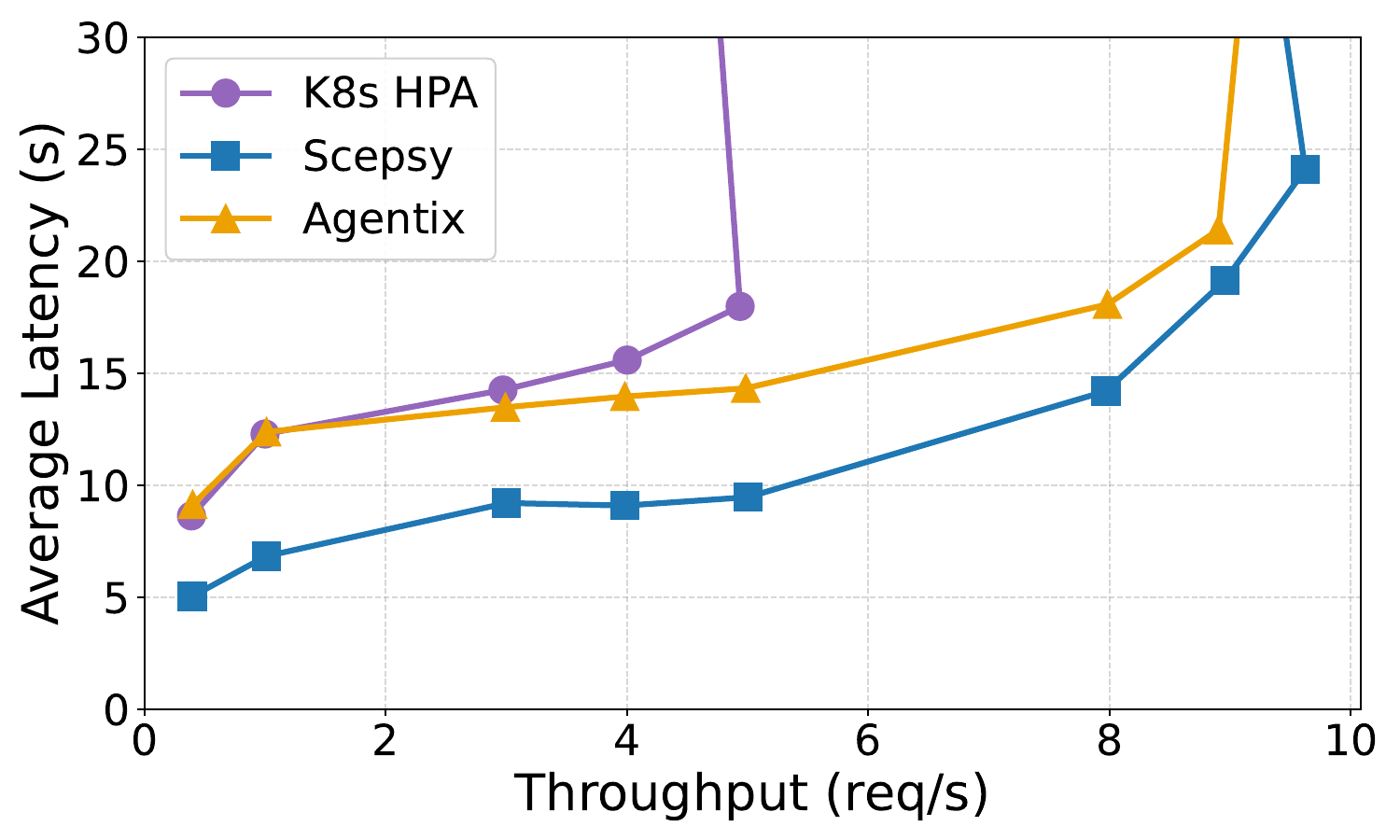}\label{fig:tl-beam-16gpu}
        \tabularnewline

        \rotatebox[origin=c]{90}{\footnotesize\textbf{Coding Agent}} &
        \centering\gridpanel{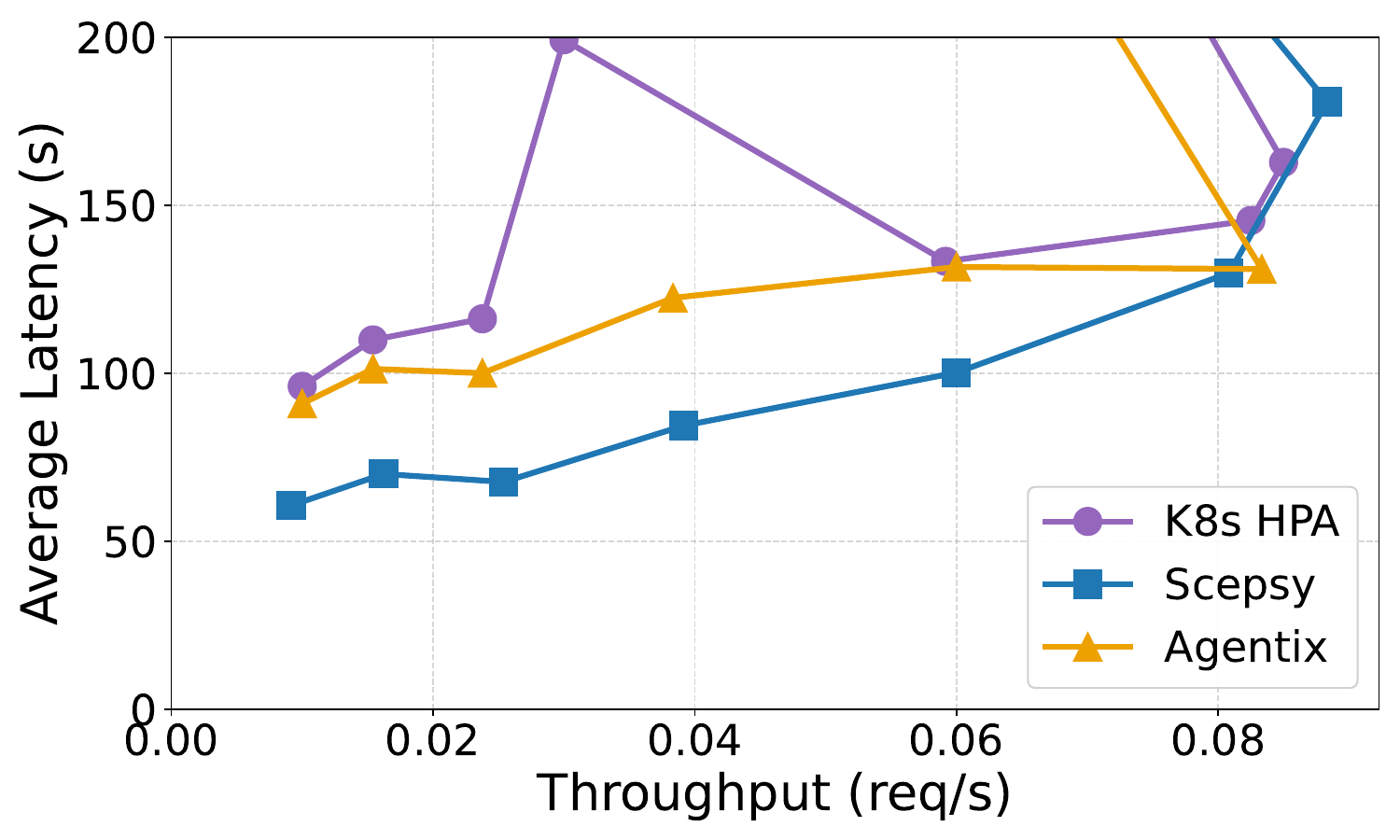}\label{fig:tl-coding-4gpu} &
        \centering\gridpanel{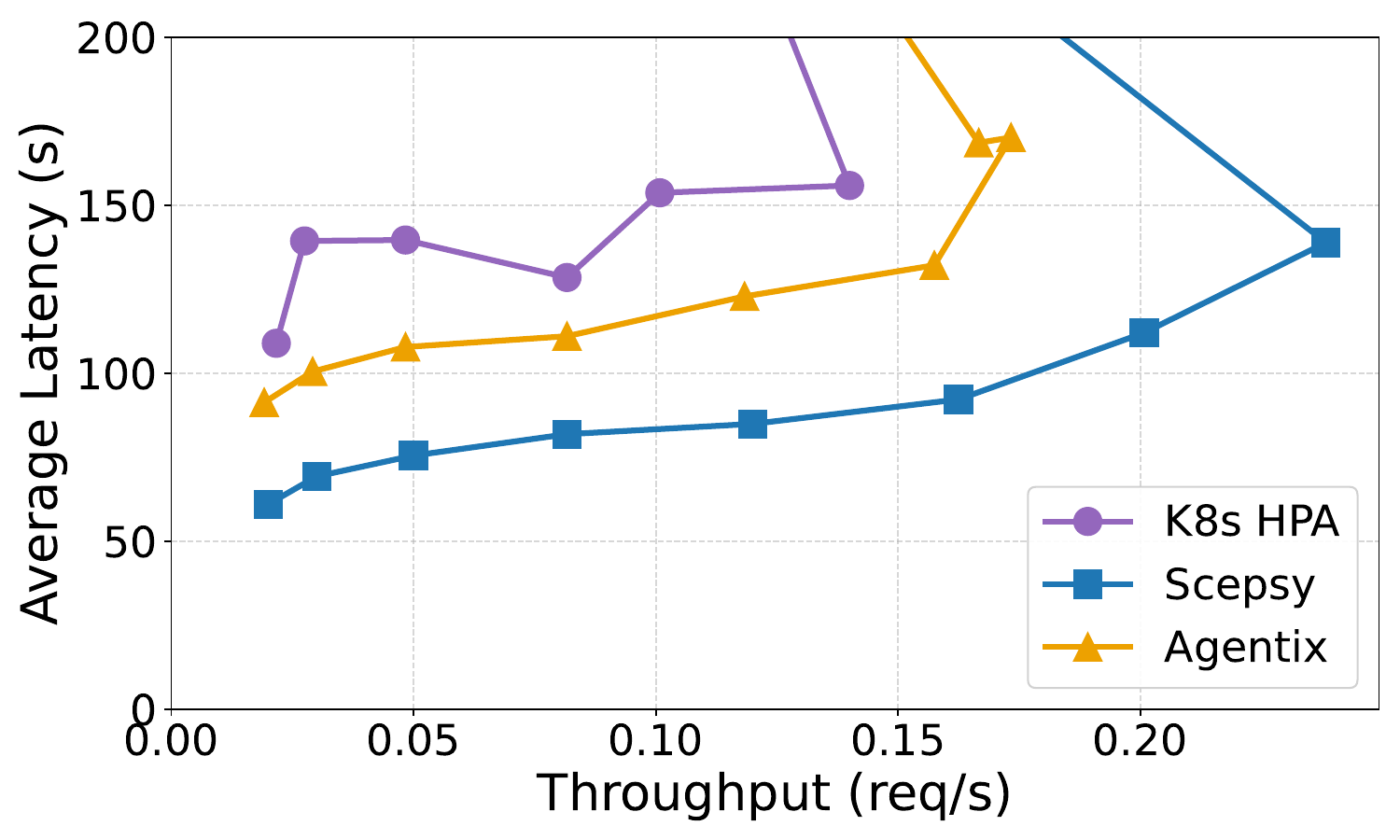}\label{fig:tl-coding-8gpu} &
        \centering\gridpanel{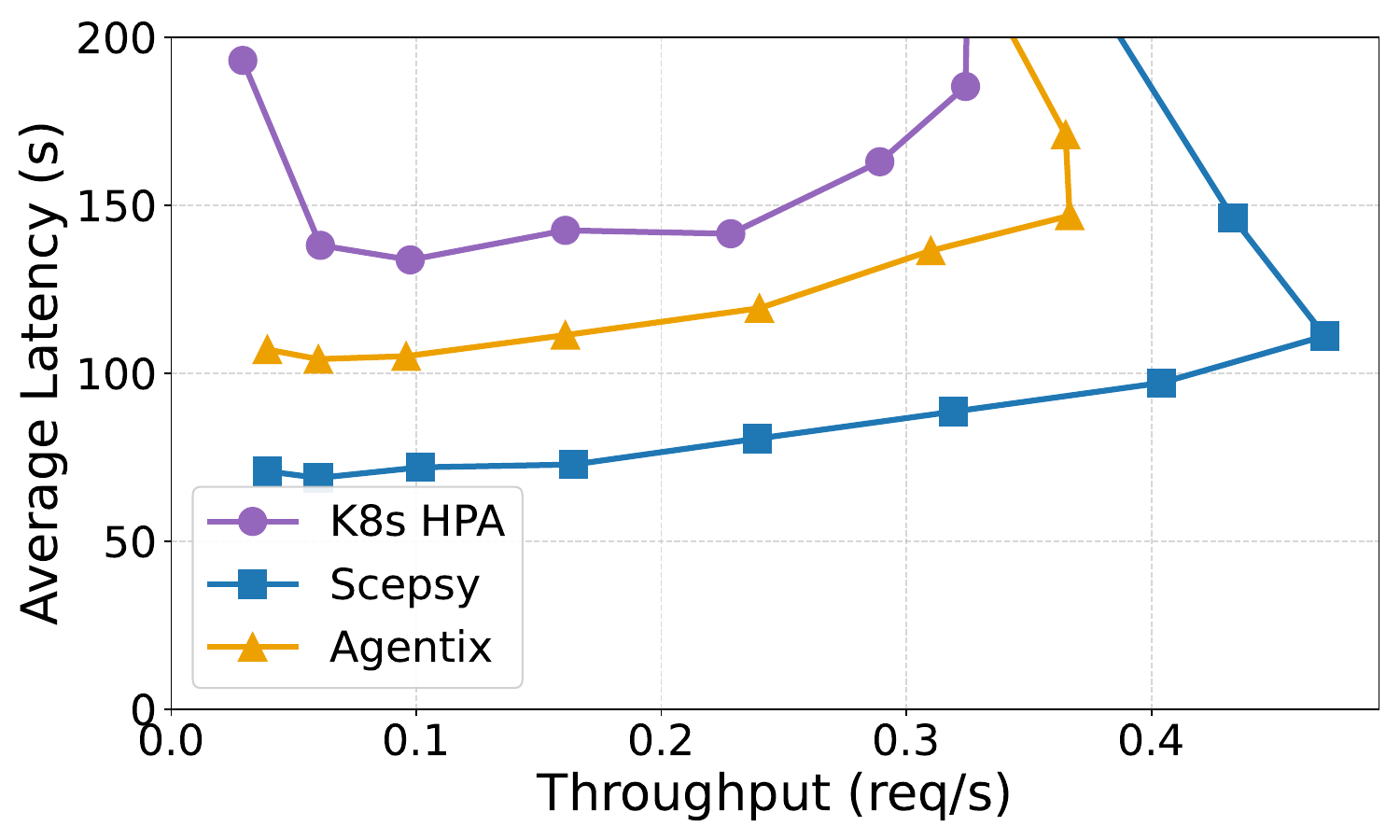}\label{fig:tl-coding-16gpu}
        \tabularnewline

    \end{tabular}
    \caption{Throughput--latency comparison across workloads and 4, 8, 16~GPUs}
    \label{fig:throughput-latency-grid}
\end{figure*}

The GPU scheduler fixes each LLM's replica count, tensor parallelism degree, and GPU share, and packs the shares onto numbered GPUs~(\Cref{sec:resource-scheduling}). Placement maps them to physical GPUs such that a tensor-parallel replica sits in one \emph{NVLink domain}, a set of GPUs connected by NVLink, and replicas that share a GPU each get a hard limit on its streaming multiprocessors~(SMs) and memory. Kubernetes' Nvidia device plugin shares a GPU only in equal slices and hands them out regardless of NVLink domain~\cite{fractopoincompat}. Placement is vector bin packing, which is NP-hard~\cite{vecbinpack}, so \sys{} places greedily, most constrained first, in two stages.

\mypar{Most-constrained-first placement in NVLink domains} Placement order decides feasibility: small replicas placed first spread over a node's GPUs and leave no NVLink domain free for a tensor-parallel replica. \sys{} therefore places the most constrained replicas first, by tensor parallelism degree and then by fraction, through Kubernetes priority classes. Within a node, our extension of the Nvidia device plugin keeps a replica of one GPU or more inside one NVLink domain and packs a smaller replica onto the fullest GPU that fits it, so whole GPUs stay free. The split within a domain mirrors the scheduler's contiguous packing: a $1.7$-GPU replica takes a whole GPU plus $0.7$ of the next, because a balanced $0.9$ and $0.8$ scatters the fractions that share its GPUs and strands the next reallocation's whole-GPU replicas.

\mypar{Enforcing fractions and restarting only changed replicas} A fraction is a hard limit, not a hint. Nvidia MPS~\cite{mps} caps each replica's SMs at its smallest per-GPU fraction, and the engine caps its memory at $90$\% of that fraction. A replica also exits if its GPUs outnumber its tensor parallelism degree or lack NVLink, so a misplacement fails rather than serves over PCIe. Reconfiguration is incremental. On a reallocation~(\Cref{sec:scheduling:adaptation}), \sys{} diffs the new placement against the running one, stops only the replicas that changed or were dropped, and places the new ones in the same order. The other replicas keep serving behind their LLM's router, an SGLang Model Gateway~\cite{sglang}, which routes by load and KV cache across them.

\section{Evaluation}

We evaluate \sys{} on three workflows: RAG\bplus{}reranker, beam search, and a coding agent. We compare against the autoscaler of Kubernetes~\cite{kubernetes}, the multi-LLM serving of Aegaeon~\cite{aegaeon}, and the workflow-aware serving of Agentix~\cite{agentix}~(\Cref{sec:e2e-workflows}), serve two workflows at once~(\Cref{sec:combined-workflow}), ablate the contributions~(\Cref{sec:ablation-study}), adapt a running allocation to workload shifts~(\Cref{sec:eval:adaptation}), and benchmark the GPU scheduling search~(\Cref{sec:scheduler-latency}).

\subsection{Experimental setup}\label{sec:eval:setup}

\mypar{Cluster} We run on-premises experiments on $16$~GPUs ($4$~nodes with $4$~GPUs each). Each node has an AMD EPYC 7402P CPU, $4\times$ Nvidia RTX A6000 GPUs, and PCIe~4.0. The nodes are connected by $100$\unit{Gbps} InfiniBand, and the GPUs of a node form two pairs over third-generation NVLink.

\mypar{Software} \sys{} runs on MicroK8s~\cite{microk8s} v1.32, with vLLM~\cite{vllm} v0.17 as the LLM engine and the Model Gateway v0.3 of SGLang~\cite{sglang} as the request router.

\mypar{Baselines} We compare \sys{} to (1)~the autoscaler of Kubernetes~\cite{kubernetes} on the same MicroK8s, (2)~Aegaeon~\cite{aegaeon, aegaeon-artifact} for multi-LLM multiplexing, and (3)~Agentix~\cite{agentix} as the workflow-aware baseline.

\mypar{Workloads} We benchmark the systems with RAG\bplus{}reranker~\cite{rag-reranker, flashrag} and beam search~\cite{search-and-learn, scaling-test-time-blog}, alone and combined. RAG\bplus{}reranker serves e5-base-v2 and Llama-3-8B, and beam search serves Llama-3.2-1B and Llama3.1-8B-PRM. The coding agent replays traces recorded with Qwen3-Coder-30B-A3B-Instruct as the main agent and Qwen2.5-7B-Instruct for the sub-agents against Llama-3.1-8B-Instruct and Llama-3.2-1B-Instruct. The replay forces the recorded token counts on every call, so the served models set the load but not the trajectory of the agent.

\subsection{End-to-end workflows}\label{sec:e2e-workflows}
We compare end-to-end throughput--latency curves, running the systems at increasing arrival rates for each workflow and cluster size. We report two numbers per comparison, both from before saturation, which starts at the first arrival rate at which throughput grows by less than $5$\% or average latency jumps by more than twice its low-load value. The throughput improvement is the ratio of the throughputs at the last unsaturated points. The latency improvement is the range of latency ratios at equal throughput over the rates at which neither system has saturated.

\mypar{Autoscaling} We compare \sys{} to the Kubernetes autoscaler~\cite{kubernetes} for all three workflows on $4$, $8$, and $16$~GPUs. On $4$~GPUs, the arrival rates span $1$--$28$~req/s for RAG\bplus{}reranker, $0.1$--$4.25$~req/s for beam search, and $0.01$--$0.28$~req/s for the coding agent. They grow with the cluster to $4$--$112$, $0.4$--$17$, and $0.04$--$0.64$~req/s on $16$~GPUs.

\Cref{fig:throughput-latency-grid} shows the achieved throughput on the x-axis and the workflow request latency on the y-axis. For all workflows and cluster sizes, \sys{} outperforms the autoscaler in throughput and latency. Across cluster sizes, \sys{} sustains $1.7$--$2.5\times$ the throughput of the autoscaler before saturation for beam search, $1.1$--$2.2\times$ for RAG\bplus{}reranker, and $1.1$--$1.7\times$ for the coding agent. The latency improvements are $1.0$--$1.9\times$, $1.2$--$3.3\times$, and $1.1$--$2.7\times$, respectively.

The autoscaler loses on both metrics because it reacts to load, whereas \sys{} predicts what each allocation yields. It scales up on queued requests, scales down once the queues are empty, and therefore oscillates between allocations. It also knows nothing of tensor parallelism, so it cannot spend a spare GPU budget on lower latency.

\begin{figure}[t]
    \centering
    \setlength{\tabcolsep}{1pt}
    \begin{tabular}{@{}c m{0.48\columnwidth} m{0.48\columnwidth}@{}}
        & \multicolumn{1}{c}{\footnotesize\textbf{4 GPUs}}
          & \multicolumn{1}{c}{\footnotesize\textbf{8 GPUs}} \\[4pt]

        \rotatebox[origin=c]{90}{\footnotesize\textbf{RAG + Reranker}} &
        \centering\includegraphics[width=\linewidth]{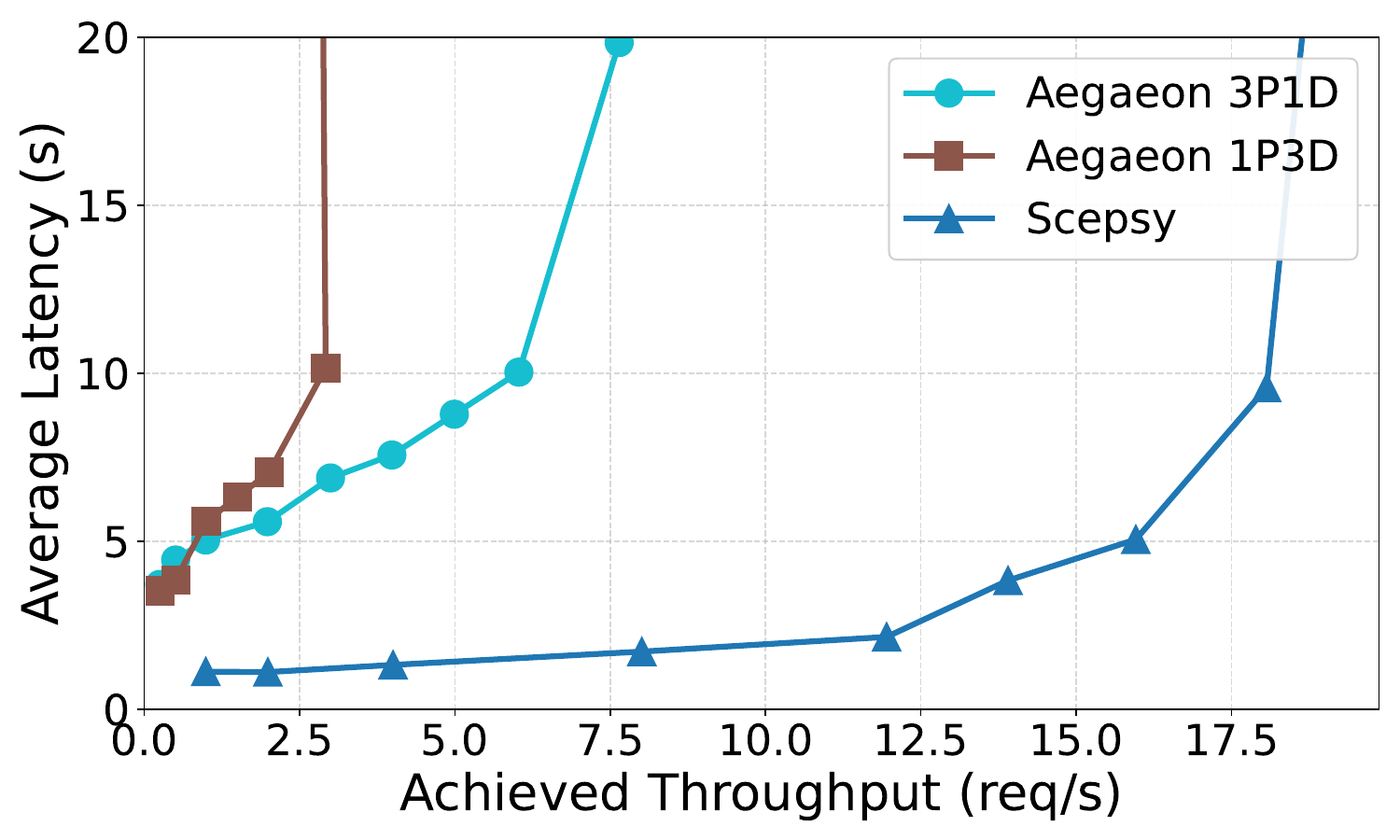}\label{fig:aegaeon-tl-rag-4gpu} &
        \centering\includegraphics[width=\linewidth]{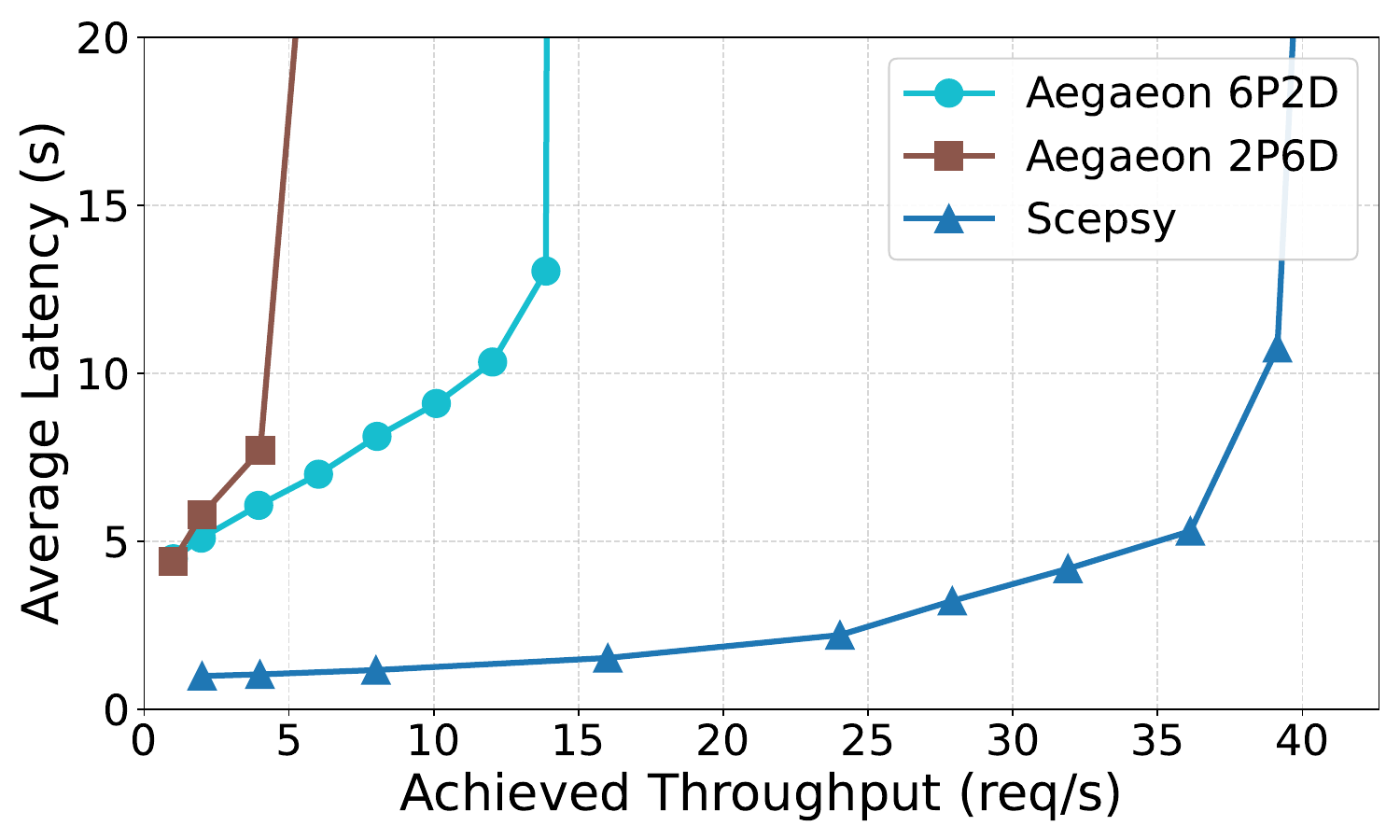}\label{fig:aegaeon-tl-rag-8gpu}
        \tabularnewline[4pt]

        \rotatebox[origin=c]{90}{\footnotesize\textbf{Beam Search}} &
        \centering\includegraphics[width=\linewidth]{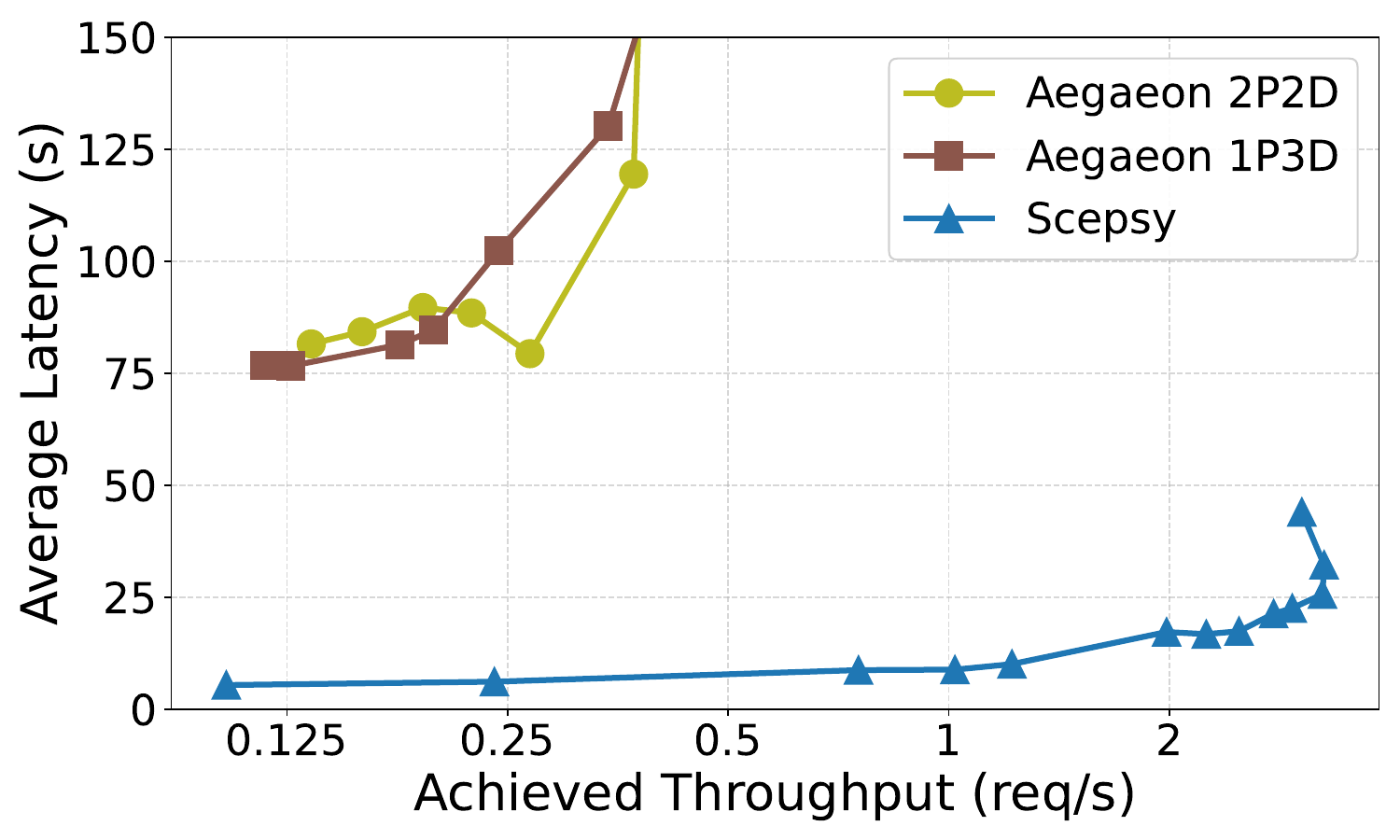}\label{fig:aegaeon-tl-beam-4gpu} &
        \centering\includegraphics[width=\linewidth]{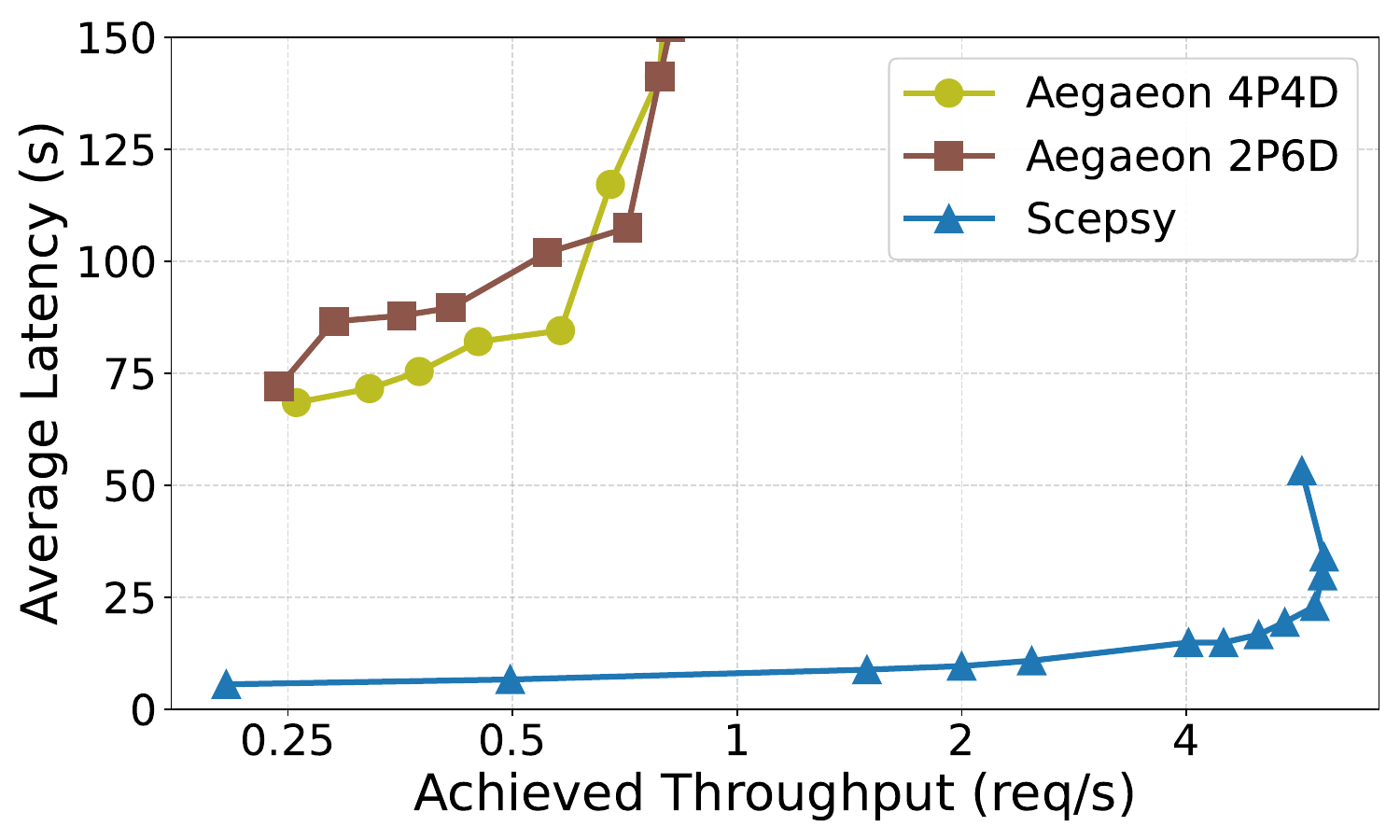}\label{fig:aegaeon-tl-beam-8gpu}
        \tabularnewline
    \end{tabular}
    \caption{Throughput--latency of \sys{} \vs Aegaeon}
    \label{fig:scepsy-vs-aegaeon-grid}
\end{figure}

\begin{figure*}[t]
    \centering
    \setlength{\tabcolsep}{1pt}
    \begin{tabular}{@{}c m{0.26\textwidth} m{0.27\textwidth} m{0.26\textwidth}@{}}
        & \multicolumn{1}{c}{\footnotesize\textbf{4 GPUs}}
          & \multicolumn{1}{c}{\footnotesize\textbf{8 GPUs}}
          & \multicolumn{1}{c}{\footnotesize\textbf{16 GPUs}} \\[2pt]

        \rotatebox[origin=c]{90}{\footnotesize\textbf{RAG + Reranker}} &
        \centering\gridpanel{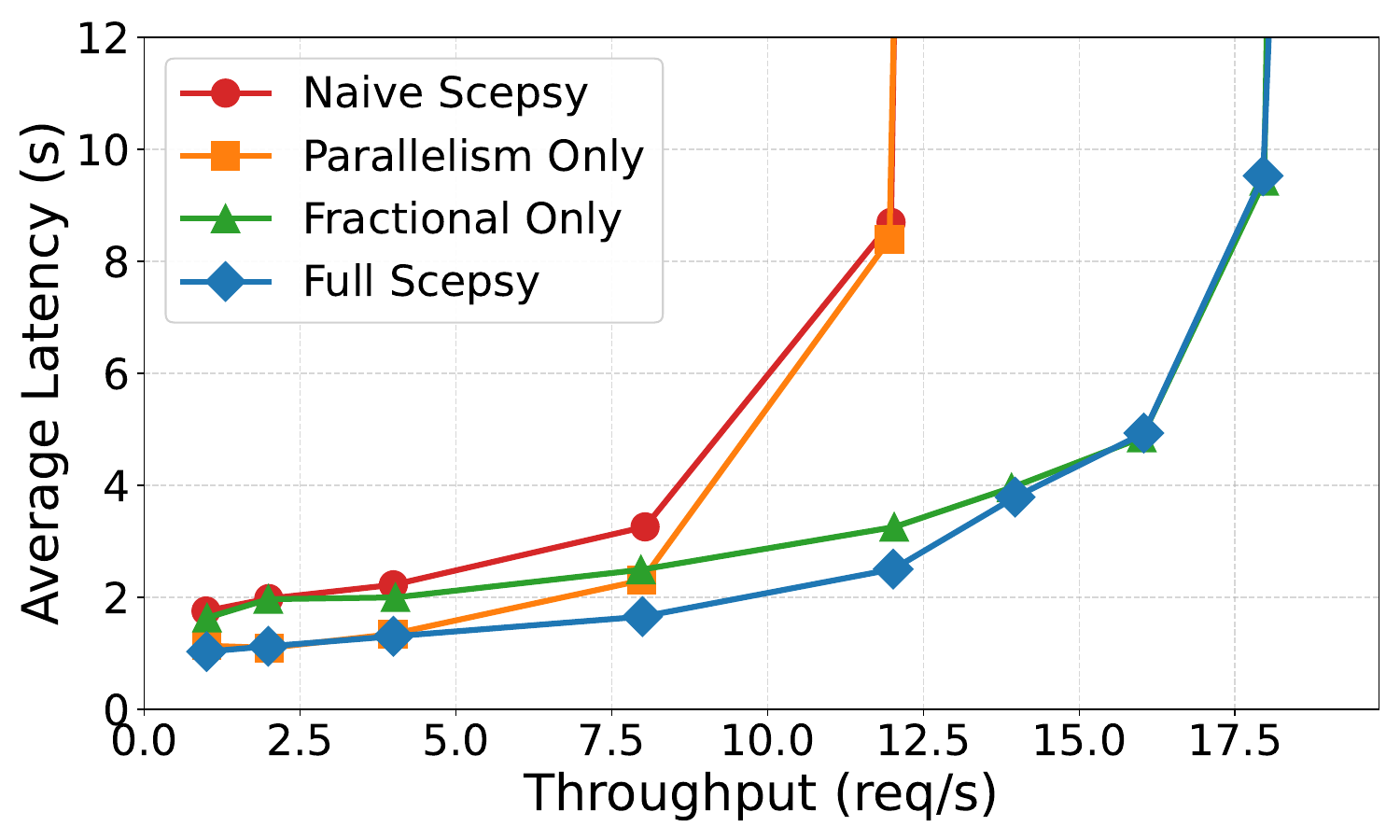}\label{fig:ablation-rag-4gpu} &
        \centering\gridpanel{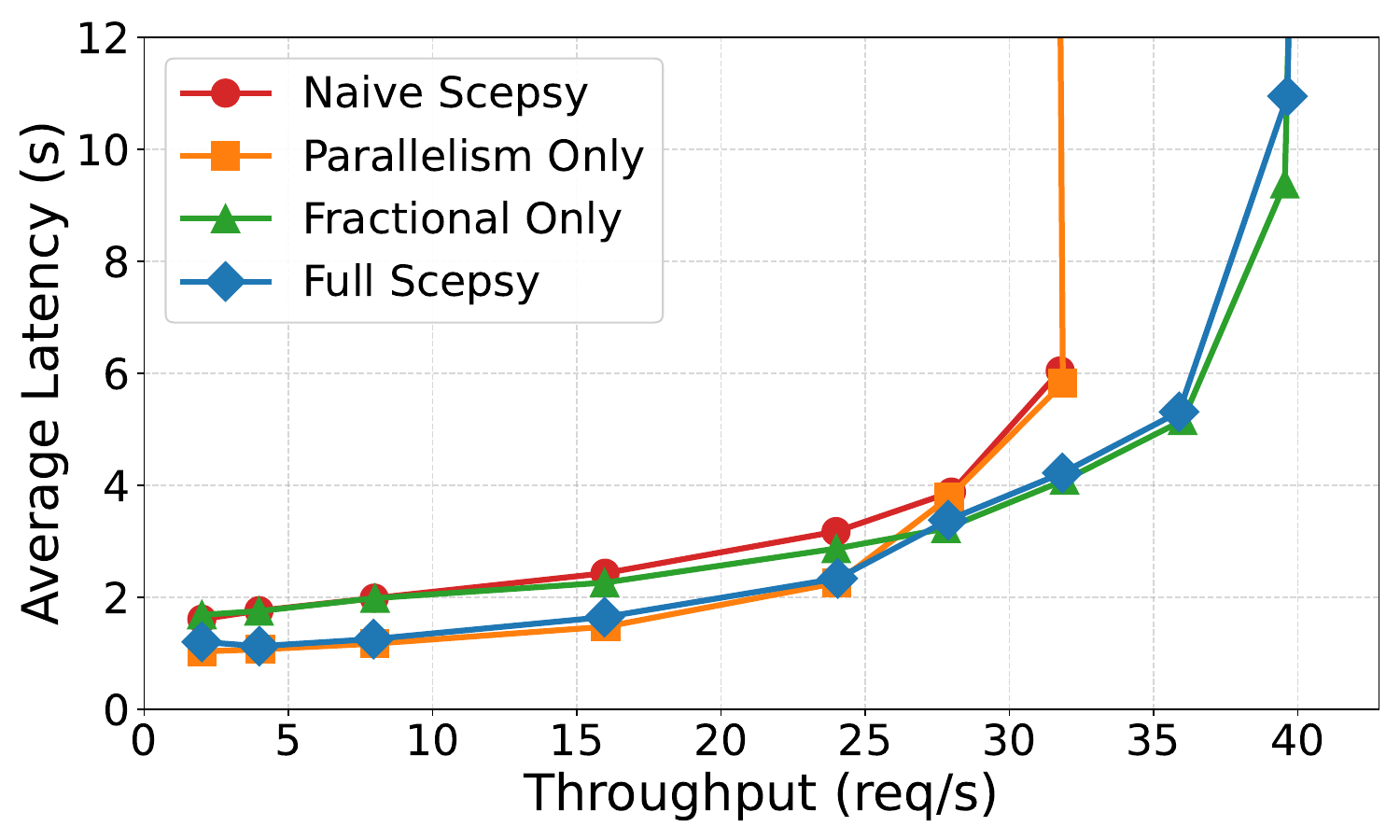}\label{fig:ablation-rag-8gpu} &
        \centering\gridpanel{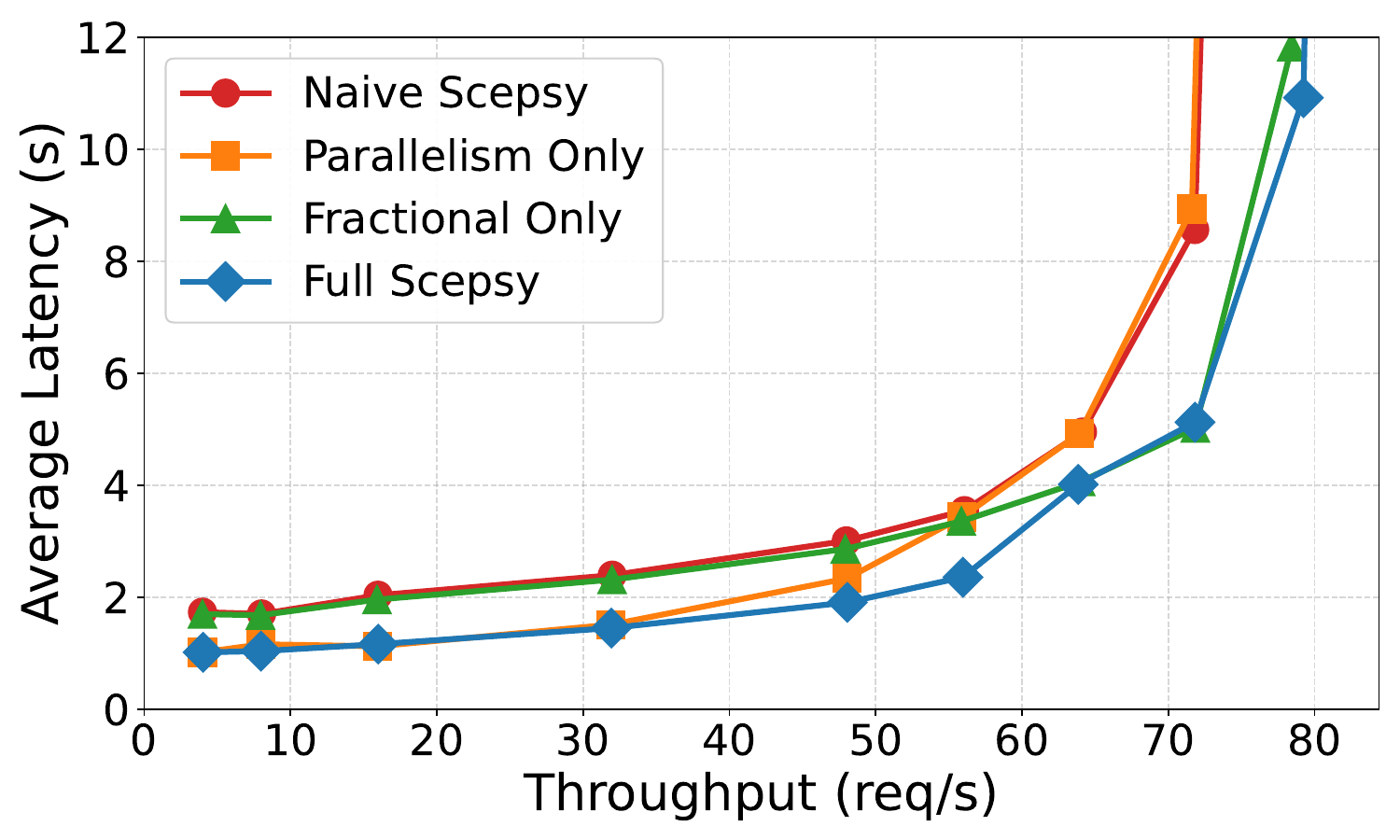}\label{fig:ablation-rag-16gpu}
        \tabularnewline

        \rotatebox[origin=c]{90}{\footnotesize\textbf{Beam Search}} &
        \centering\gridpanel{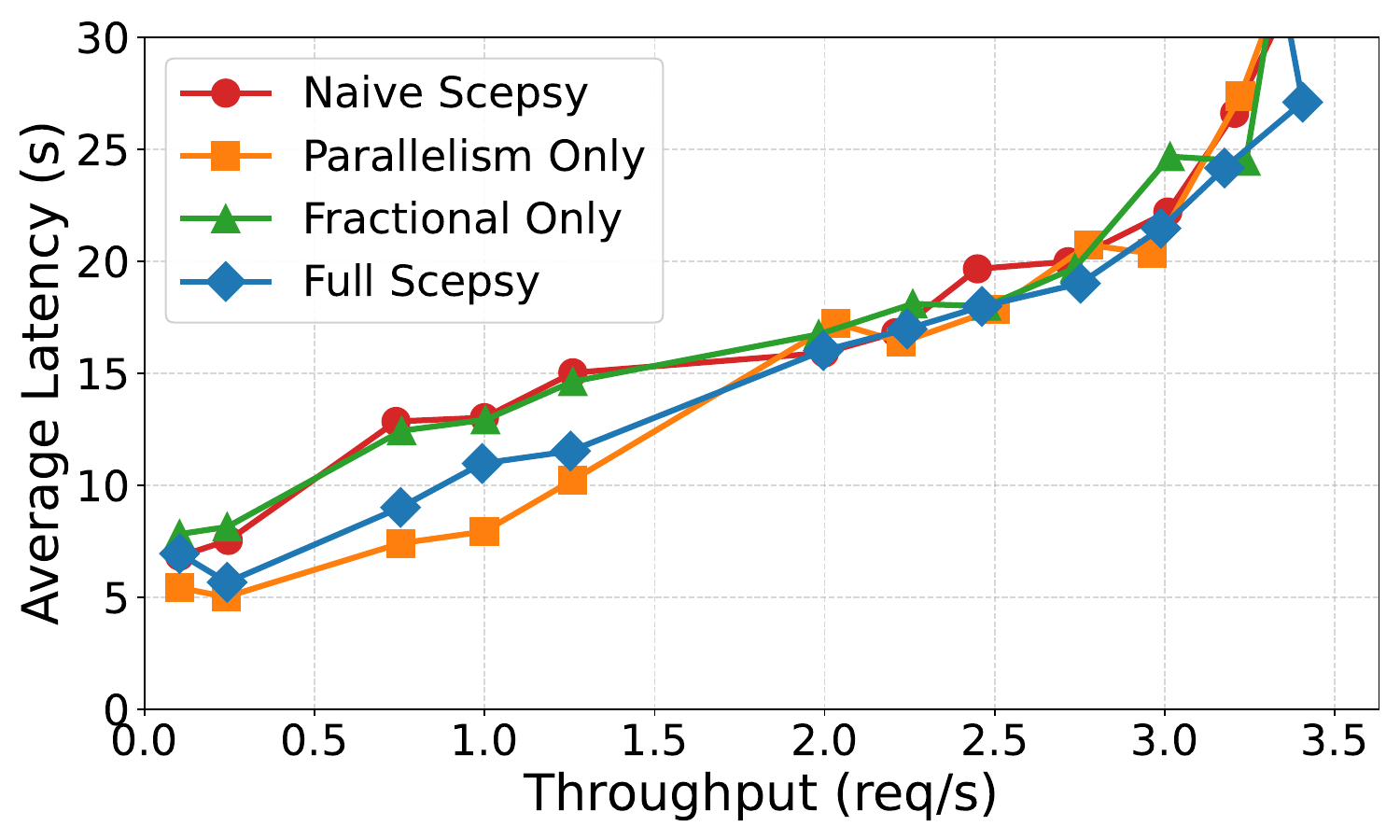}\label{fig:ablation-beam-4gpu} &
        \centering\gridpanel{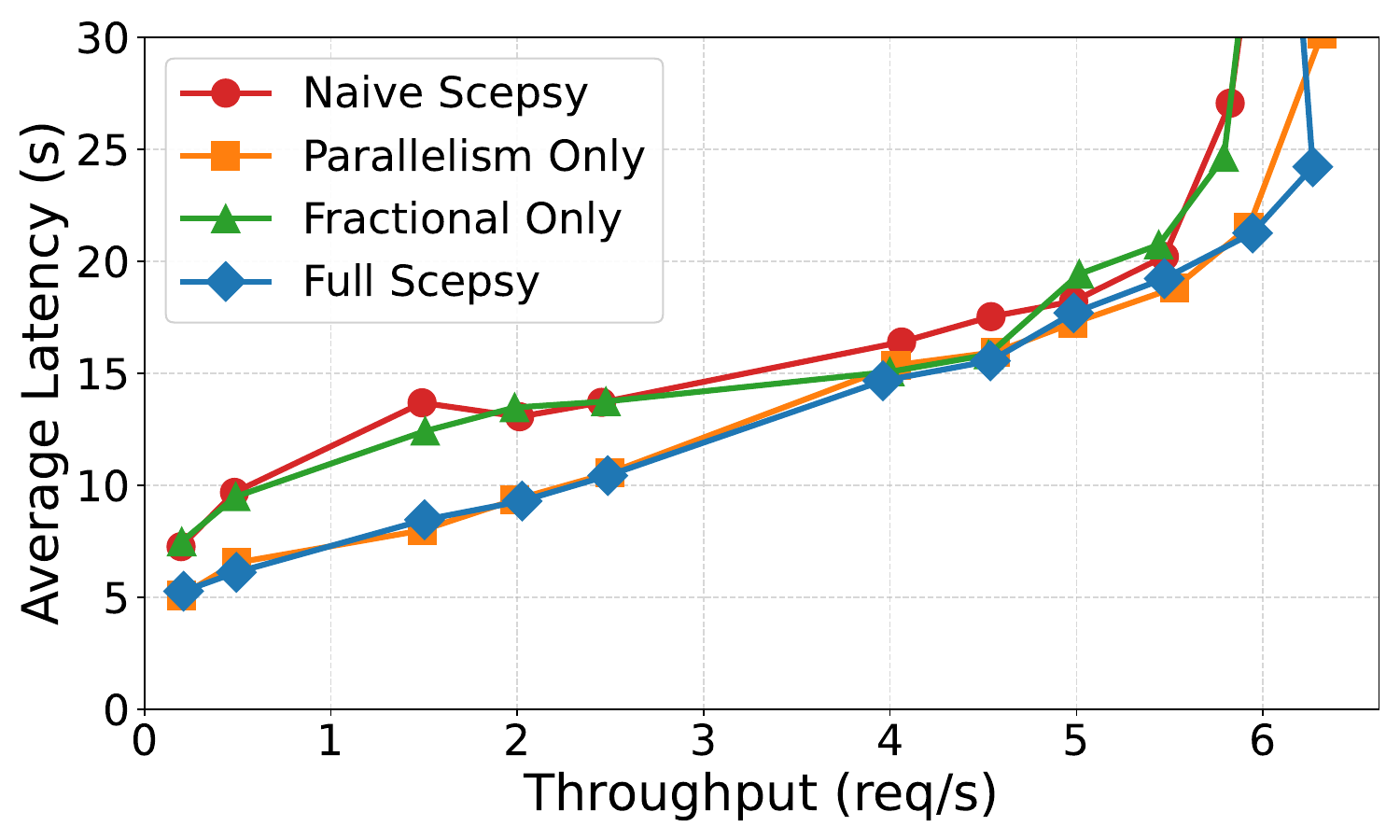}\label{fig:ablation-beam-8gpu} &
        \centering\gridpanel{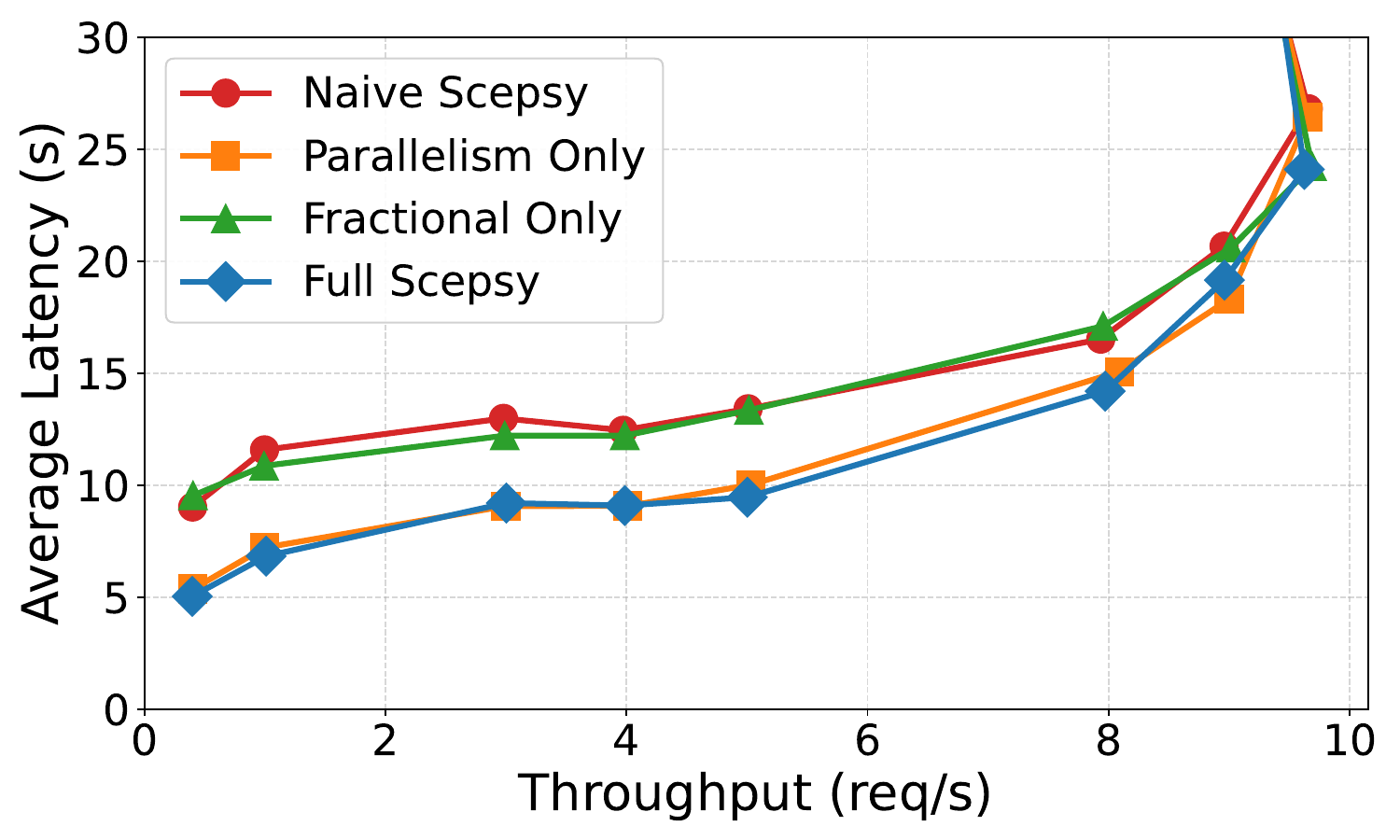}\label{fig:ablation-beam-16gpu}
        \tabularnewline

        \rotatebox[origin=c]{90}{\footnotesize\textbf{Coding Agent}} &
        \centering\gridpanel{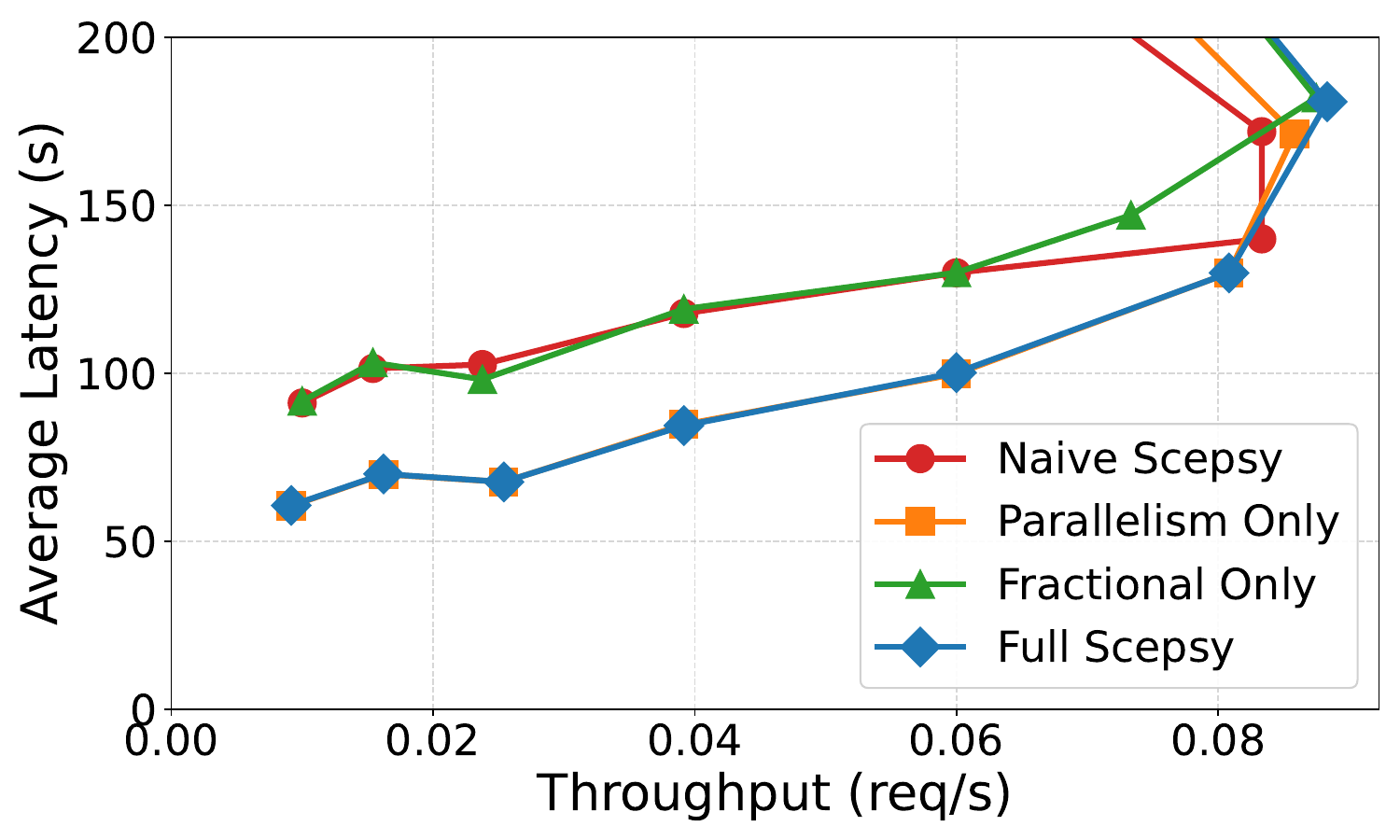}\label{fig:ablation-coding-4gpu} &
        \centering\gridpanel{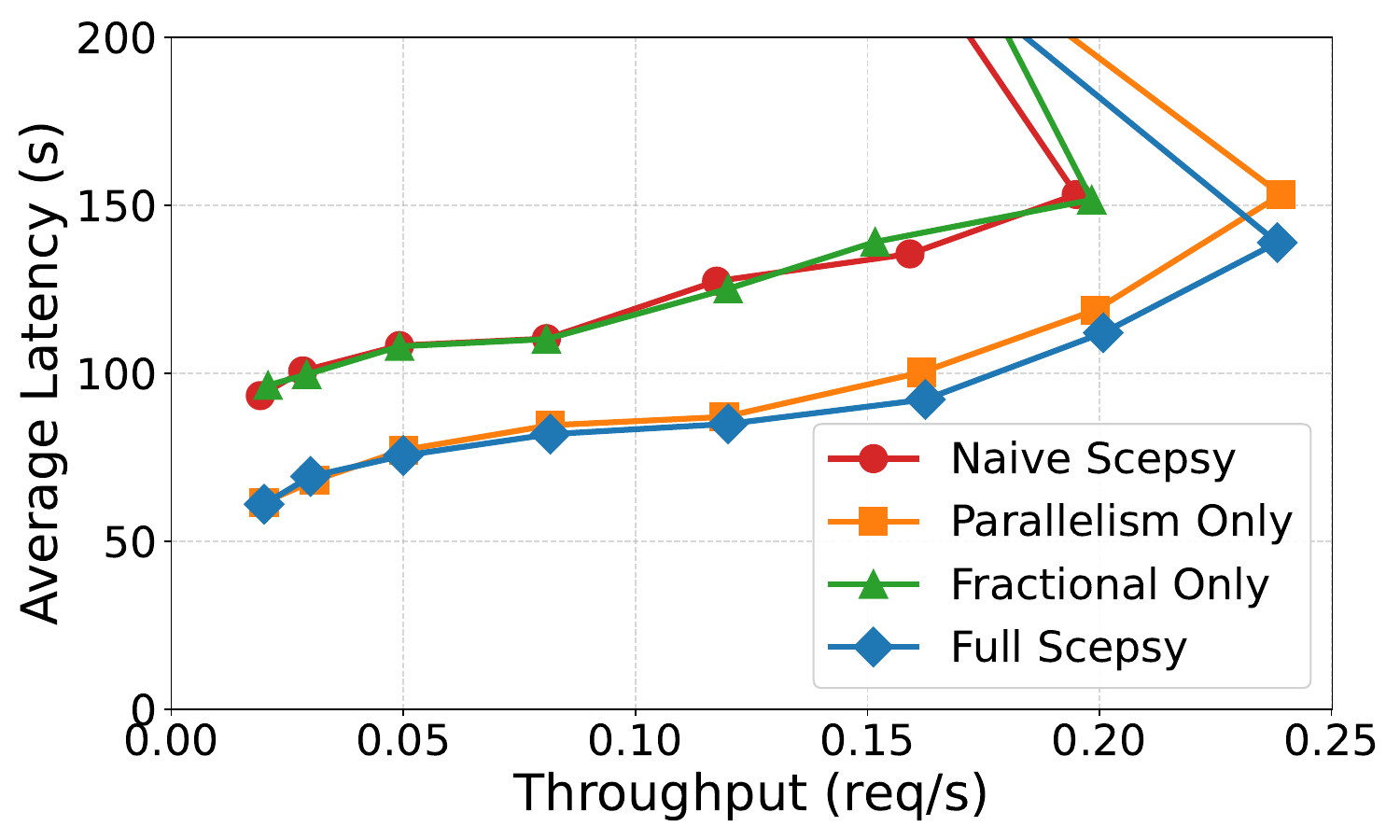}\label{fig:ablation-coding-8gpu} &
        \centering\gridpanel{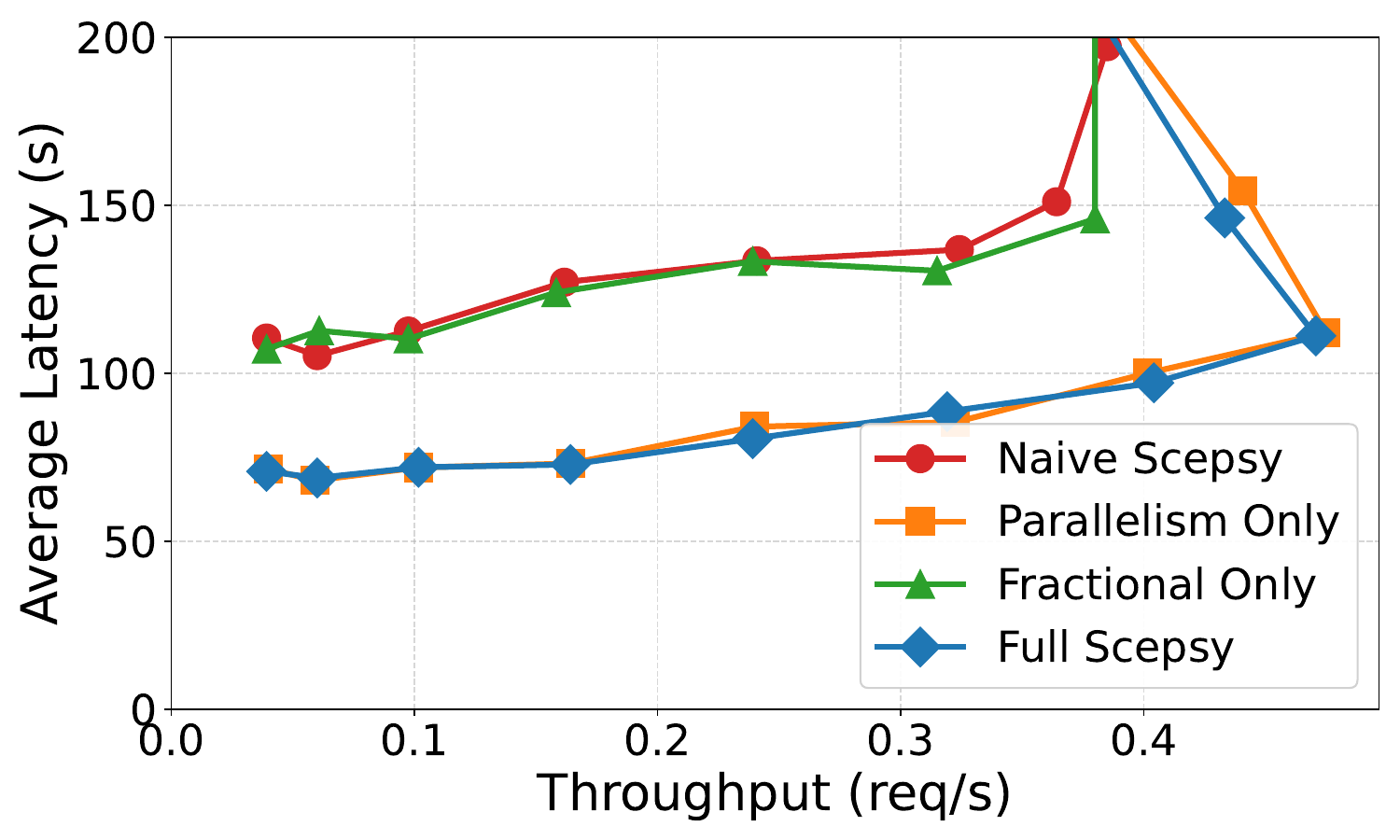}\label{fig:ablation-coding-16gpu}
        \tabularnewline

    \end{tabular}
    \caption{Ablation study showing the contribution of \sys{}'s key optimizations across workloads and 4, 8, 16~GPUs}
    \label{fig:ablation-study}
\end{figure*}

\mypar{LLM multiplexing} We compare \sys{} against Aegaeon~\cite{aegaeon}, a multi-model serving system that pools GPUs across models through token-level autoscaling. Aegaeon disaggregates serving into prefill and decode instances, and we serve the non-LLM models (reranker, embedding) from its prefill instances. We test three per-node prefill/decode splits, 3P/1D, 2P/2D, and 1P/3D, and use the same per-node split on both nodes for $8$~GPUs.

\Cref{fig:scepsy-vs-aegaeon-grid} shows RAG\bplus{}reranker and beam search on $4$ and $8$~GPUs. \sys{} outperforms every Aegaeon split for both workloads. Before saturation, \sys{} sustains $7.5$--$8.7\times$ the throughput of the best split for beam search and $2.6\times$ for RAG\bplus{}reranker, with latency improvements of $11.9$--$19.6\times$ and $3.2$--$9.1\times$ over the plotted splits.

Aegaeon lacks prefix caching, which penalizes beam search most because its candidate solutions share long prefixes, and it pays a swap of prefill instances per model switch, whereas \sys{} keeps every LLM resident.

\mypar{Agentic workflow} We run the workflow-aware baseline, Agentix~\cite{agentix}, on vLLM v0.17. Instead of the custom swap kernel of the original paper, we swap KV cache blocks through the offloading connector of vLLM. We cap the offloading space at $8$\unit{GiB}, because a large swap space would favor Agentix for a feature that \sys{} could enable as well. Agentix runs on a fixed GPU allocation, the one under which \sys{} reaches its maximum throughput.

\Cref{fig:throughput-latency-grid} also shows Agentix. \sys{} sustains $1.1\times$ the throughput of Agentix for beam search on all cluster sizes, $1.1$--$2.0\times$ for RAG\bplus{}reranker, and $1.1$--$1.4\times$ for the coding agent. The latency improvements are $1.0$--$1.8\times$, $1.2$--$2.0\times$, and $1.0$--$1.5\times$, respectively.

For beam search, the gain of \sys{} comes from choosing a different parallelism configuration. Both systems saturate at a similar throughput, because the throughput-optimal allocation for beam search uses no fractional GPUs. For RAG\bplus{}reranker, \sys{} gains from both the parallelism configuration and fractional GPU allocation. For the coding agent, the parallelism configuration alone lowers latency. Agentix is workflow-aware but leaves the parallelism configuration unchosen, and that choice is \sys{}'s gain.

\subsection{Combined workflows}\label{sec:combined-workflow}

\begin{figure}[t]
    \centering
    \begin{subfigure}[b]{0.48\columnwidth}
        \centering
        \includegraphics[width=\textwidth]{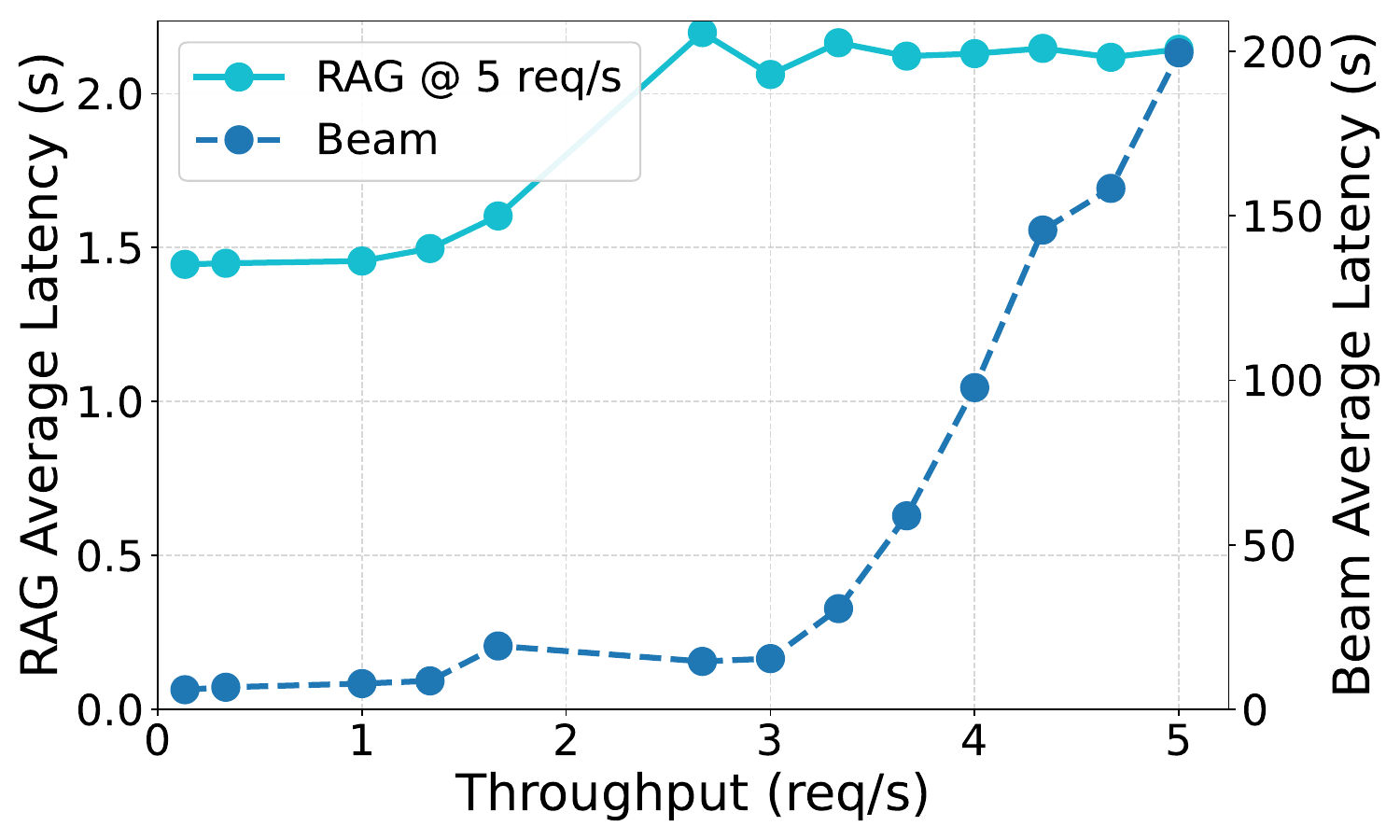}
        \caption{RAG\bplus{}reranker held}
        \label{fig:combined-effect-rag}
    \end{subfigure}
    \hfill
    \begin{subfigure}[b]{0.48\columnwidth}
        \centering
        \includegraphics[width=\textwidth]{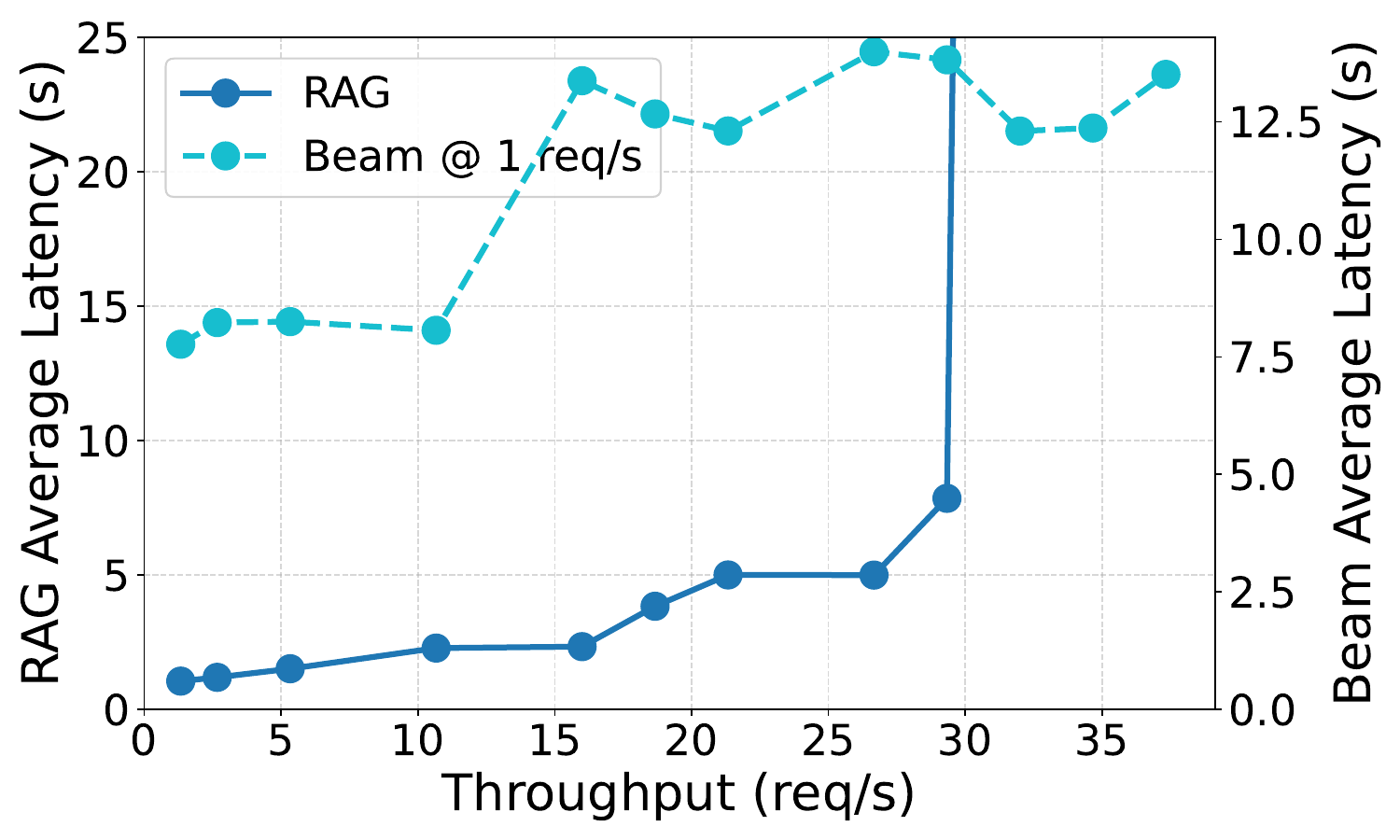}
        \caption{Beam search held}
        \label{fig:combined-effect-beam-search}
    \end{subfigure}
    \caption{Effect of combined workflows on 8~GPUs}
    \label{fig:combined-effect}
\end{figure}

To show that \sys{} adapts the allocation to the arrival rates of several workflows, we fix the rate of one workflow and sweep that of the other on $8$~GPUs. RAG\bplus{}reranker holds $5$~req/s while beam search sweeps~(\Cref{fig:combined-effect-rag}), and beam search holds $1$~req/s while RAG\bplus{}reranker sweeps~(\Cref{fig:combined-effect-beam-search}). As the swept workflow fills the cluster, its latency rises until it saturates, whereas the held workflow keeps its throughput and its latency rises by less than $2\times$. One allocation thus serves both workflows without starving either, because the scheduler splits the GPUs under the egalitarian policy, which maximizes the utility of the worst-served workflow~(\Cref{sec:scheduling:search}).

\subsection{Performance impact of contributions}\label{sec:ablation-study}

To understand the contribution of \sys{}'s key optimizations, we evaluate the throughput--latency curve with features disabled. We run four configurations: the full system, without tensor parallelism (fractional only), without co-location (parallelism only), and with both disabled, for all three workflows on $4$, $8$, and $16$~GPUs. \Cref{fig:ablation-study} shows the curves. For RAG\bplus{}reranker on all cluster sizes, co-location is the biggest contributor to the throughput improvement. For beam search on all cluster sizes, tensor parallelism accounts for the latency reduction.

The coding agent benefits from tensor parallelism alone, and co-location changes nothing. The two configurations with tensor parallelism run $26$--$54$\unit{s} faster ($1.3$--$1.7\times$), up to $0.06$~req/s on $4$~GPUs and up to the knee on $8$ and $16$~GPUs, where they also raise the throughput before saturation by $21$--$25$\%. With both optimizations disabled, \sys{} loses latency and throughput at once. Which optimization dominates thus depends on the workflow: co-location for RAG\bplus{}reranker, tensor parallelism for beam search and the coding agent.

\begin{figure*}[t]
    \centering
    \begin{minipage}[b]{0.235\textwidth}
        \centering
        \resizebox{\linewidth}{!}{\input{figures/adaptation/beam_search_width_avg_8gpus.pgf}}
        \caption{Adaptation to a beam-width shift from 4 to 16}
        \label{fig:adaptation-width}
    \end{minipage}
    \hfill
    \nextfloat
    \begin{minipage}[b]{0.74\textwidth}
        \centering
        \begin{subfigure}[b]{0.31\linewidth}
            \centering
            \resizebox{\linewidth}{!}{\input{figures/search_time_vs_num_models.pgf}}
        \end{subfigure}
        \hfill
        \begin{subfigure}[b]{0.31\linewidth}
            \centering
            \resizebox{\linewidth}{!}{\input{figures/search_time_vs_total_gpus.pgf}}
        \end{subfigure}
        \hfill
        \begin{subfigure}[b]{0.31\linewidth}
            \centering
            \resizebox{\linewidth}{!}{\input{figures/search_time_vs_num_fractions.pgf}}
        \end{subfigure}
        \caption{Search time as scheduling parameters scale for the combined workflow}
        \label{fig:search-time}
    \end{minipage}
\end{figure*}

\subsection{Adaptation to workload shifts}\label{sec:eval:adaptation}

So far, each run used one allocation computed for a fixed arrival rate. Next, we shift the workload during a run: it holds for $5$\unit{min} (RAG\bplus{}reranker) or $10$\unit{min} (beam search), shifts for $20$\unit{min}, and shifts back for $20$\unit{min}. The plots show throughput and average latency in $30$\unit{s} bins, the detection of the shift (dotted), and the instant at which the new allocation is in place (solid).

\mypar{Arrival-rate shifts} RAG\bplus{}reranker steps from $6$ to $16$~req/s ($4$~GPUs) and from $12$ to $28$~req/s ($8$~GPUs), and beam search from $0.75$ to $2.8$ and from $1$ to $3.75$~req/s~(\Cref{fig:adaptation-single}). \sys{} detects each shift within $29$--$61$\unit{s} and has the new allocation in place within $138$--$183$\unit{s}. At the up-shift, the initial allocation cannot serve the higher rate, and RAG\bplus{}reranker latency climbs to $18$--$20$\unit{s}. The scheduler trades tensor parallelism for replicas, \eg three single-GPU replicas of Llama-3-8B instead of one tensor-parallel replica on $4$~GPUs. Throughput then matches the offered rate, and latency settles at $4$--$6$\unit{s} for RAG\bplus{}reranker and $16$--$23$\unit{s} for beam search. At the down-shift, \sys{} restores the latency-optimal allocation, and latency returns to its baseline. Each reallocation shows as a throughput dip of about $100$\unit{s} while the redeployed LLM restarts.

\mypar{Combined workflows} RAG\bplus{}reranker and beam search run together, and the rate of one workflow steps up $2$--$3\times$ while the other holds constant~(\Cref{fig:adaptation-combined}). Detection takes $23$--$56$\unit{s}, and the new allocation is in place within $144$--$234$\unit{s}. The constant workflow keeps its throughput, because only the LLMs of the shifted workflow are redeployed, but its latency rises as the scheduler rebalances the cluster toward the shifted workflow: on $8$~GPUs, beam search latency rises from $8$ to $13$\unit{s} while RAG\bplus{}reranker runs at $18$~req/s. After the shift back, both workflows return to their baselines.

\mypar{Workflow-structure shifts} Next, the structure of the workflow shifts rather than its load. The arrival rate holds at $0.5$~req/s ($4$~GPUs) and $1$~req/s ($8$~GPUs), and instead the beam width of every request steps from $4$ to $16$ and back~(\Cref{fig:adaptation-width}), which quadruples the LLM invocations per request at an unchanged request rate. \sys{} detects the shift from the observed call pattern alone, $124$--$268$\unit{s} after the up-shift and $92$--$107$\unit{s} after the shift back. Detection is slower than for rate shifts because the invocation statistics come from completed requests. \sys{} splits the generator LLM into twice as many single-GPU replicas, in place $235$--$369$\unit{s} after the up-shift, and reverts it $203$--$218$\unit{s} after the shift back. Latency averages $20$--$22$\unit{s} in the wide phase, about $2.7\times$ the baseline for $4\times$ the LLM work per request, and returns to $8$\unit{s} afterward.

Overall, \sys{} adapts to shifts in load and in workflow structure within $138$--$369$\unit{s} and reallocates exactly once per shift, unlike the oscillating Kubernetes autoscaler~(\Cref{sec:e2e-workflows}), because it acts on the predicted gain of a new allocation rather than on a load signal, at the cost of a reconfiguration window of about $100$\unit{s} per reallocation.

\begin{figure}[t]
    \centering
    \setlength{\tabcolsep}{1pt}
    \begin{tabular}{@{}m{0.48\columnwidth} m{0.48\columnwidth}@{}}
        \multicolumn{1}{c}{\footnotesize\textbf{RAG + Reranker}}
          & \multicolumn{1}{c}{\footnotesize\textbf{Beam Search}} \\[4pt]
        \centering\resizebox{\linewidth}{!}{\input{figures/adaptation/rag_reranker_avg_8gpus.pgf}}\label{fig:adaptation-single-rag-8gpu} &
        \centering\resizebox{\linewidth}{!}{\input{figures/adaptation/beam_search_avg_8gpus.pgf}}\label{fig:adaptation-single-beam-8gpu}
        \tabularnewline
    \end{tabular}
    \caption{Adaptation to rate shifts, single workflows on 8~GPUs}
    \label{fig:adaptation-single}
\end{figure}

\begin{figure}[t]
    \centering
    \setlength{\tabcolsep}{1pt}
    \begin{tabular}{@{}m{0.48\columnwidth} m{0.48\columnwidth}@{}}
        \multicolumn{1}{c}{\footnotesize\textbf{RAG shift}}
          & \multicolumn{1}{c}{\footnotesize\textbf{Beam shift}} \\[4pt]
        \centering\resizebox{\linewidth}{!}{\input{figures/adaptation/combined_avg_rag_8gpus.pgf}}\label{fig:adaptation-combined-rag-8gpu} &
        \centering\resizebox{\linewidth}{!}{\input{figures/adaptation/combined_avg_beam_8gpus.pgf}}\label{fig:adaptation-combined-beam-8gpu}
        \tabularnewline
    \end{tabular}
    \caption{Adaptation to rate shifts, combined workflow on 8~GPUs}
    \label{fig:adaptation-combined}
\end{figure}

\subsection{Scheduler latency}\label{sec:scheduler-latency}

The space of GPU allocations is large, so we measure the search time of the GPU scheduler for the combined workflow while varying the number of LLMs, GPUs, and fractions per GPU~(\Cref{fig:search-time}).

With $16$~GPUs and $10$~fractions per GPU, the search time grows exponentially with the number of LLMs but stays below $35$\unit{s} (left). With $3$~LLMs and $10$~fractions per GPU, it also grows exponentially with the number of GPUs and reaches $70$\unit{s} at $128$~GPUs (middle). With $3$~LLMs and $16$~GPUs, it reaches $1$\unit{s} at $20$~fractions per GPU (right). The search is thus not the bottleneck of a reallocation: even at $128$~GPUs it takes $70$\unit{s}, less than the $100$\unit{s} restart of one redeployed replica~(\Cref{sec:eval:adaptation}).

\section{Related Work}
\label{sec:related-work}

\myparr{LLM serving engines} optimize inference for a single LLM\@. Within a replica, engines page the KV cache~(vLLM~\cite{vllm}), batch chunked prefill with decode~\cite{sarathi-serve, deepspeedfastgen}, overlap operations with complementary resource demands~(NanoFlow~\cite{nanoflow}), and fuse prefill and decode attention~(POD-Attention~\cite{podattention}). Across replicas, they disaggregate prefill from decode~\cite{distserve, mooncake}, switch the parallelism configuration between the two phases~\cite{loongserve, seesaw}, evict parameters to make room for the KV cache~(KunServe~\cite{kunserve}), scale a multimodal model's components independently~(CornServe~\cite{cornserve}), or cut the time to load a model onto a GPU~\cite{serverlessllm, torpor, blitzscale}. None of them sees the workflow that invokes the LLM\@. \sys{} is an orchestration layer above them and drives any engine through its completions API\@.

\myparr{GPU sharing} multiplexes a GPU across jobs in time or in space~\cite{tgs, bullet, sirius, seallm}. Temporal multiplexing time-slices the GPU across models, as Nvidia vGPU~\cite{vgpu} and AlpaServe~\cite{alpaserve} do, so only one model runs at a time and the others wait even when its kernels leave SMs idle. Spatial multiplexing runs them concurrently: through a per-process SM cap~(Nvidia MPS~\cite{mps}), by co-scheduling kernels whose compute and memory demands do not interfere~(Orion~\cite{orion}), or by padding real-time kernels with best-effort kernels~(Reef~\cite{reef}). \sys{} uses Nvidia MPS to co-locate LLM replicas and to cap each replica's SMs at its GPU fraction~(\Cref{sec:implementation}).

\section{Conclusion}

\sys{} serves multi-LLM agentic workflows from any agentic framework on a GPU cluster. Workflow latency is unpredictable, but each LLM's fraction of the execution time is stable across requests. \sys{} combines these fractions with per-LLM profiles into an \alp{} that predicts workflow throughput and latency for an allocation. It uses the \alp{} to search jointly over fractional GPU shares, tensor parallelism degrees, and replica counts across a workflow's LLMs. A hierarchical heuristic places the allocation onto the cluster, respecting its topology. On three workflows, \sys{} sustains up to $2.5\times$ higher throughput before saturation and $1.0$--$3.3\times$ lower latency than the Kubernetes autoscaler.

\bibliographystyle{plain}
\bibliography{scepsy.bib}

\end{document}